\RequirePackage{fix-cm}
\documentclass[twocolumn]{svjour3}          
\smartqed  
\usepackage{graphicx}
\usepackage{mathptmx}
\usepackage{cite}
\usepackage{hyperref}
\hypersetup{
    colorlinks = true,
    citecolor = blue,
    linkcolor = red,
}
\usepackage{filecontents,lipsum}
\usepackage{subcaption}
\usepackage[skip=2pt,font=scriptsize]{caption}
\usepackage[skip=2pt,font=scriptsize]{subcaption}
\usepackage{amsmath,amssymb,amsfonts}
\usepackage{textcomp}
\usepackage{color}
\usepackage[dvipsnames]{xcolor}
\begin{document}

\title{Exploring Chaos and Ergodic behavior of an Inductorless Circuit driven by Stochastic Parameters 
\thanks{S. Seth and A. Bera have contributed equally.}
}

\titlerunning{Exploring Chaos and Ergodic behavior of an Inductorless Chaos Generator Circuit driven by stochastic parameters}        

\author{Soumyajit Seth \and Abhijit Bera \and Vikram Pakrashi}
\authorrunning{Seth et al.} 
\institute{
S. Seth* \and V. Pakrashi \at UCD Centre for Mechanics, Dynamical Systems and Risk Laboratory, School of Mechanical and Materials Engineering, University College Dublin, Ireland. \\
\email{soumyajitseth01@gmail.com} \\
* Corresponding author \\
V. Pakrashi \\
\email{vikram.pakrashi@ucd.ie}
\and
A. Bera \at Department of Physical Sciences, Indian Institute of Science Education and Research Kolkata, Mohanpur Campus, 741246, IN \\
\email{abhijitbera95@gmail.com}
}

\date{Received: date / Accepted: date}

\maketitle

\begin{abstract}

There exist extensive studies on periodic and random perturbations of various continuous maps investigating their dynamics. This paper presents a random piecewise smooth map derived from a simple inductor-less switching circuit. The bifurcation parameter is bounded and randomly selected from a stationary distribution. Due to the stochasticity inherent in either the parameter values or the state variable, the time evolution of the state variable cannot be predicted at a specific time instant. We observe that the state variable exhibits completely ergodic behavior when the minimum value of the parameter is $2.0$. However, the ensemble average of the state variable converges to a fixed value. For parameter values ranging from $2.0$ to $3.5$, the system demonstrates nonchaotic behavior and the absolute value of the Lyapunov exponent increases monotonically with the asymmetry ($a_p$) of the distribution from which the bifurcation parameter values are sampled. We determine the probability density function of the random map and verify its invariance under any initial condition. The most noteworthy result is the disappearance of chaotic behavior when the lower range of the distribution is varied while maintaining a fixed upper threshold for a particular distribution, even though the nonrandom map exhibits an array of periodic and chaotic behaviors within that range. 


\end{abstract}

\section{Introduction}
\label{intro}
In the last four decades, it has been found that a large class of physical and engineering systems, such as power electronics circuits, impacting systems, stick-slip oscillations, walking and hopping mechanisms in robotics, etc., can be portrayed as switching dynamical systems. Under stroboscopic sampling, these non-autonomous switching dynamical systems may give rise to piecewise smooth maps \cite{4766286, 703837, PhysRevE.79.037201}. Such systems exhibit a special class of dynamical phenomena called `border collision bifurcations' \cite{PhysRevE.59.4052, 847870, 841921}, where under the variation of one of the parameters, the  location of fixed point changes. At a critical parameter value, it collides with the border. Such an event abruptly changes the stability status of the map, resulting in a rich array of bifurcation phenomena, which do not occur in smooth dynamical systems.

Period incrementing cascade with chaotic inclusions is a particular sequence of border collision bifurcations that occur in one-dimensional switching dynamical systems \cite{Avrutin_2006, ParthaSharathiDutta}. As a system parameter varies smoothly in one direction, the system goes through a series of border collision bifurcations resulting in a sequence of periodic windows in-between chaotic attractors. The periodicity in each window is one greater than that of the previous one \cite{Avrutin_2006, PhysRevE.75.066205}.

A simple one-dimensional, non-autonomous, inductorless switching electronic circuit is considered in a number of papers \cite{1333221, seth2016study} in this regard. This circuit is significant in practical applications, especially in chaotic communications because it has no inductor components and can generate robust chaos \cite{PhysRevLett.80.3049} phenomena under specific parameter settings. It has been experimentally demonstrated that this circuit shows the period incrementing cascade with chaotic inclusion bifurcation phenomenon while the input DC voltage, used as a bifurcation parameter, varies smoothly in a reverse direction \cite{iiserkeprints1110}.

Stochasticity is inherent in every dynamical phenomenon in reality for any physical system, leading to uncertainties in the state variables caused by different reasons, such as fluctuations in parameter spaces, switching surfaces, initial conditions, etc. Consequently, based on theoretical and numerical analyses, the dynamics of a system exhibit deviations from what is expected from its deterministic behavior.

Previous works related to stochastic continuous maps \cite{Chamayou1991, Bhattacharya1993, Steinsaltz1999}, have explored whether the state variables tend to converge to zero over long-term iterations, thereby exhibiting a delta function of state variables at zero during significant time intervals. It has been confirmed that, under certain conditions, this delta function at zero-valued state variables indeed exists.

Several works have also focused on studying the dynamics of piecewise smooth maps in the presence of stochasticity. A quantitative analysis conducted by Simpson and Kuske \cite{simpson2012stochastically} investigated the effects of small-amplitude, additive, white Gaussian noise on stable sliding motion for a piecewise smooth map. Additionally, in another study by Simpson and Kuske \cite{Simpson2015}, they examined the stochastic dynamics near a periodic orbit when a small noise was embedded in a general piecewise-smooth vector field, providing a theoretical perspective on the subject.

Furthermore, Simpson et al. \cite{doi:10.1137/120884286} conducted numerical investigations on a specific type of two-dimensional piecewise smooth map known as the Nordmark map. They introduced additive Gaussian noise with a fixed amplitude and observed irregular behaviors resulting from the noise, causing transitions between different dynamical behaviors as the bifurcation parameter varied. Rounak and Gupta \cite{Rounak2020} showed numerically the effect of randomness in the forcing function on a harmonically excited bilinear impact oscillator with a soft barrier, which lead to qualitatively different dynamical behaviors, such as various periodic and chaotic attractors at different parameter values due to the presence of multiple coexisting attractors.

It has been numerically shown that the normal form of piecewise smooth maps, under the periodic or stochastic variations of the functional form, can lead to significant changes in their dynamics. These changes arise due to the coexistence of state- and time-dependent switchings \cite{Mandal_2016, MANDAL2017532, MANDAL2018154} or the stochastic variation of the system's borders \cite{MANDAL20172161, mandal2019dynamics}. The functional form of the map of the switching system changes either periodically or stochastically due to the presence of both kinds of switching in a system. Under the periodic variation of the functional form of the one-dimension map in any one compartment of the phase space, the bifurcation from a stable fixed point attractor to a periodic attractor of a period greater than one due to the border collision bifurcation has been established. When the functional form changes stochastically, the dynamical behavior of any two orbits may differ even if they start from the same initial point inside the non-deterministic basin of attraction. On the other hand, the piecewise smooth map with a stochastically varying border, varying stochastically inside a small region of the phase space, gives rise to a non-deterministic basin of attraction. Similar to the case of stochastically varying functional forms, the dynamical behavior of any two orbits, starting from the same initial point inside this non-deterministic attraction basin may differ.

It is observed from the existing literature that in previous investigations, stochasticity has been incorporated either through the functional forms of piecewise-smooth maps or by introducing randomness at the borders of the map. In this study, we have adopted a different approach by introducing stochasticity in the parameter space of the map chosen from a probability distribution within a specific parameter range. By doing so, we aim to explore whether the system exhibits any new dynamical behaviors other than the dynamics obtained from the non-random map when varying the distribution. Additionally, this approach allows for investigating and uncovering any characteristics, such as the interplay of periodic and chaotic orbits, that may reveal the presence of the discrete deterministic non-random map in the parameter regime. To demonstrate these aspects, a one-dimensional piecewise smooth map derived from an inductorless chaos generator circuit is chosen, which was also implemented in electronic experiments \cite{iiserkeprints1110}. Unlike previous approaches, where the input voltage parameter was considered to have a fixed value, this paper considers it a random variable independently chosen from a distribution and can capture the realistic variations in the system and the consequences of such realistic variations in terms of their dynamical responses. 

\section{System under investigation}
\label{ckt_sys}

\begin{figure}[tbh]
\centering
\includegraphics[width=0.7\columnwidth]{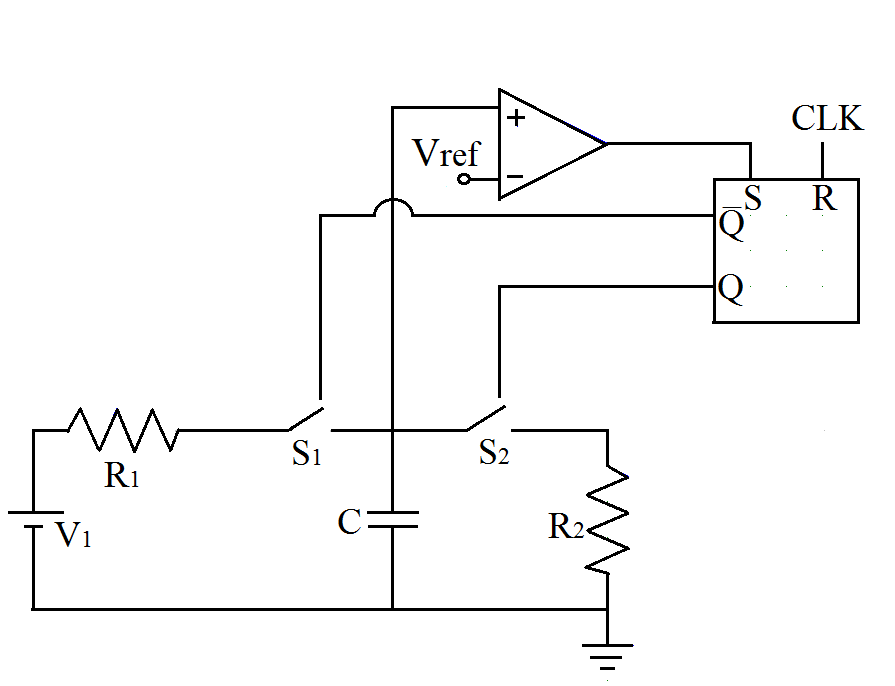}
\caption{The switching circuit under investigation}
\label{ckt}
\end{figure}
Fig.~\ref{ckt} illustrates a simple inductorless chaos generator circuit suitable for IC implementation  \cite{1333221, seth2016study}.

The circuit consists of a capacitor $C$, which can be charged through a DC input voltage $V_1$ and resistance $R_1$ or discharged through a resistance $R_2$. Two switches, $S_1$ and $S_2$, control the charging and discharging processes. These switches are controlled by a set-reset ($S$-$R$) flip-flop. A clock signal (CLK) with a period $T$ resets the flip-flop, initiating the charging phase of the capacitor $C$. While the voltage across the capacitor, denoted as $v_C$, is below a reference voltage $V_{\rm ref}$, $S_1$ remains ON and $S_2$ OFF. During this period, any clock pulse that arrives does not have any effect. Once the voltage $v_C$ reaches $V_{\rm ref}$, the latch sets, turning $S_1$ OFF and $S_2$ ON, and the capacitor starts discharging through $R_2$. The arrival of the next clock pulse resets the latch, and the charging mode is turned on again.

The evolution of the state variable $v_C$ from one clock instant to the next can occur in two possible ways. Either $v_C$ reaches the reference voltage before the arrival of the next clock pulse, leading to an ON period followed by an OFF period, or the clock pulse arrives before $v_C$ reaches $V_{\rm ref}$, resulting in a continuous ON period lasting for $T$ time. The type of evolution that occurs during a specific clock period depends on the rate of charging, which, in turn, depends on the input voltage $V_1$. Thus, $V_1$ is considered the bifurcation parameter in this context. The clock is practically generated by producing a periodic frequency waveform $f = \frac{1}{T}$ with a very low duty cycle. The parameters $C$, $R_1$, $R_2$, $V_{\rm ref}$, and $f$ are kept fixed in the analysis. Assuming all components to be ideal, the system's governing equations are given by:
\begin{equation}
\frac{dv_{\rm C}(t)}{dt} = \left\{\begin{array}{lrl}
        \frac{V_1 - v_{\rm C}(t)}{CR_{\rm 1}}, & \text{for }& v_{\rm C}(t) < V_{\rm ref}\\
          -\frac{v_{\rm C}(t)}{CR_{\rm 2}}, & & {\rm otherwise}
        \end{array}\right.
        \label{1D_eq}
\end{equation}

\begin{figure}[tbh]
\centering
\begin{subfigure}[b]{0.9\linewidth}
\includegraphics[width=\linewidth]{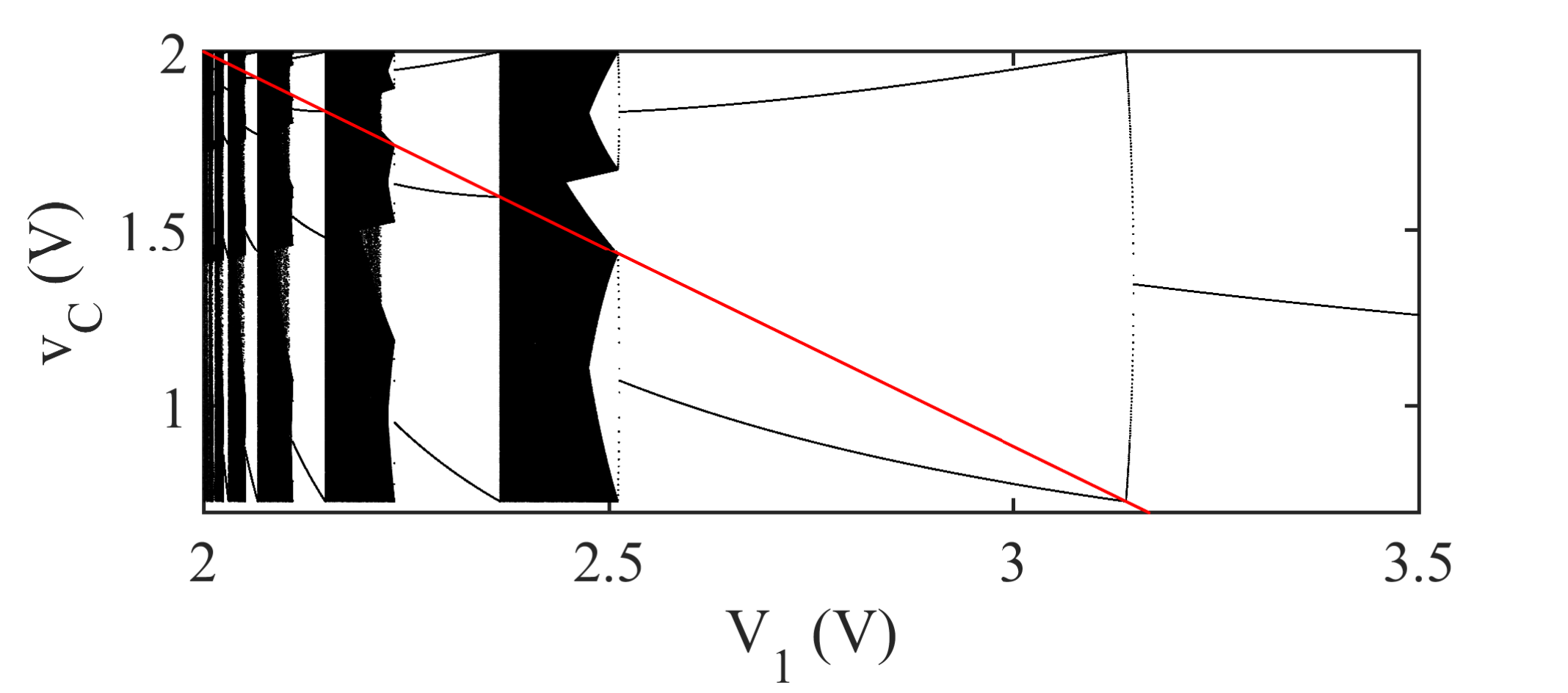}
\caption{}
\end{subfigure}
\begin{subfigure}[b]{0.8\linewidth}
\includegraphics[width=\linewidth]{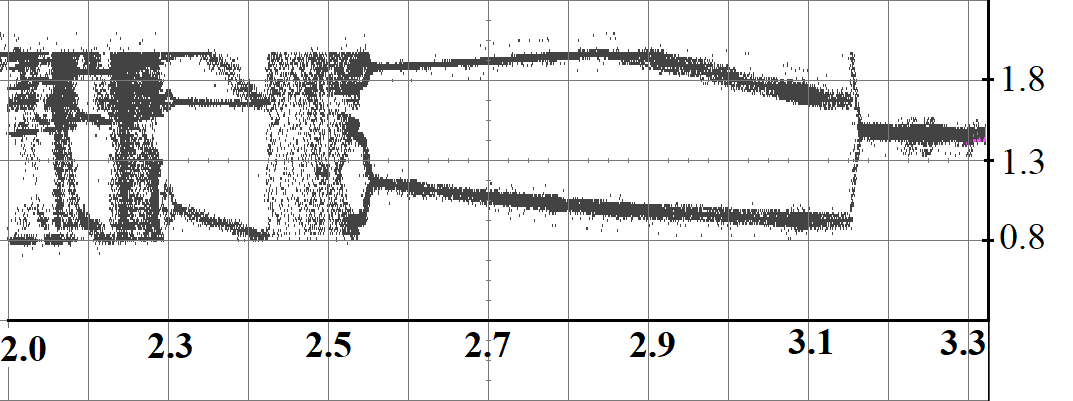}
\caption{}
\end{subfigure}
\caption{(a) Numerically and (b) Experimentally obtained bifurcation diagrams of the system~\ref{ckt}. The $x$-axis is the value of the bifurcation parameter $V_1$, and the $y$-axis is the sampled value of the capacitor voltage, $v_{\rm C}(t)$. The red line in the numerically obtained figure denotes the border voltage, $V_b$, in the parameter space $V_1$. The parameter values are: $R_1 = 9.0~\mathrm{k \Omega}, R_2 = 6.7~\mathrm{k \Omega}, C = 9.9~\mathrm{nF}, f = 15~\mathrm{kHz}$, and $V_{\rm ref} = 2.0~\mathrm{V}$. (Color inline.)}
\label{map_bif}
\end{figure}
After deriving the stroboscopic map of the system using~(\ref{1D_eq}), which relates the values of the state variable $v_C(t)$ at a clock instant with that at the previous one, we obtain:
\begin{equation}\label{eq_1}
	V_{\mathrm{n}+1}=\left\{\begin{array}{lll}
		V_1-\left(V_1-V_{\mathrm{n}}\right) \cdot e^{-\frac{T}{C R_1}}, & \text { for } & V_{\mathrm{n}} \leq V_{\mathrm{b}} \\
		V_{\mathrm{ref}}\left(\frac{V_1-V_{\mathrm{n}}}{V_1-V_{\text {ref }}}\right)^{\frac{R_1}{R_2}} e^{-\frac{T}{C R_2}}, & \text { for } & V_{\mathrm{n}} \geq V_{\mathrm{b}}
	\end{array}\right.
\end{equation}
where, $V_n$ and $V_{\rm n+1}$ are the values of the voltage across the capacitor $C$ at the $n$-th and $(n+1)$-th clock instants, respectively. The border is defined as $V_{\text{b}} = V_1 - \left(V_1-V_{\text{ref}}\right) e^{\frac{T}{C R_1}}$. The numerically and experimentally obtained bifurcation diagrams are shown in Fig.~\ref{map_bif}(a) and Fig.~\ref{map_bif}(b), respectively \cite{iiserkeprints1110}. The constant parameter values are provided in the caption of Fig.~\ref{ckt}. These bifurcation diagrams reveal the reverse period-incriminating cascade phenomena with chaotic windows.

In \cite{KHALEQUE2014599}, the dynamics of a random logistic map was shown where the parameter is chosen from a distribution. The logistic map is continuous and shows periodic windows between chaotic attractors in the bifurcation diagram. The period-incriminating cascade bifurcation phenomenon only occurs in switching dynamical systems under the variation of one of the parameter values \cite{HALSE2003953, doi:10.1142/S0218127403008533, Avrutin_2006}, where periodic windows with finite widths between chaotic attractors are obtained. The difference lies in the fact that the periodic windows have some particular identities as compared to the periodic windows in the continuous map. Therefore, the study is around the dynamics of the system with the introduction of stochasticity within the parameter of the piecewise smooth map.


\section{Methods of introducing stochasticity}
\label{stochasticity}

In realistic conditions, most dynamical phenomena, whether characterized by smooth or piecewise smooth dynamics, exhibit an inherent stochasticity stemming from randomness in initial conditions of the state variables or fluctuations in parameter values. Additionally, in the case of nonsmooth dynamical systems, stochasticity arises due to fluctuations in border values. In this study, we have utilized the system~(\ref{ckt}) as a fundamental model to investigate its dynamics in the presence of stochasticity within the parameter space. Two distinct probability distributions are employed to introduce stochasticity: uniform and triangular.

To visualize the piecewise smooth map described in Equation~(\ref{eq_1}), sequential observations of the state variable, $v_C$, were conducted, synchronized with the time period of an external clock pulse. The values of the state variable at the $n$-th and $(n+1)$-th periods of the clock signal were recorded as $V_n$ and $V_{\rm n+1}$, respectively. Instead of observing the entire waveform of the state variable, the focus was on discrete points corresponding to the increasing edge of the pulse. This relationship can be generally represented as follows:
\begin{equation}
V_{\rm n+1} = f\left(V_{\rm n}\right)
\label{eq_2}
\end{equation}

The general representation of the map~(\ref{eq_2}) mentioned above is deterministic, which means that at any instant, we can characterize by fixed parameter values. The DC input voltage, $V_1$, is successively varied for the purpose of observing the qualitative change in the dynamics. In order to introduce stochasticity into the system, a modification to Equation~(\ref{eq_2}) is proposed as follows:
\begin{equation}
V_{\rm n+1}(t+1) = g\left(V_{\rm n},V_1(n)\right)
\label{eq_3}
\end{equation}

While the functional form of the map remains unchanged as described in Equation (\ref{eq_2}), the parameter values at each iteration instant, denoted as $V_1(n)$, are now stochastically selected from a predefined distribution within the bounds of $q_1$ and $q_2$. This updated version is referred to as a `stochastic map.'

In this context, a pertinent question arises concerning the potential manifestation of behavior of the nonrandom map within this probabilistic mapping, such as whether there exists any telltale sign of chaos or periodicity like the deterministic version of the map. Also, how the dynamics of the nonrandom map changes when incorporating the stochasticity in the parameter space. Moreover, we aim to ascertain whether this probabilistic system exhibits ergodicity and, if so, how it distinguishes itself from its deterministic counterpart. To address these inquiries, two methods for computing ensemble averages are adopted, which serve as tools to explore the implications of inherent stochasticity \cite{PhysRevE88040101, KHALEQUE2014599, doi:10.1142/S0129183115500862}:
\begin{itemize}
\item The Traditional Method (TM).
\item The Nature versus Nurture Method (NVN).
\end{itemize}

\subsection{TM method}
\label{tm}
In this approach, the process is initiated by fixing a small value, denoted as $\Delta_{t}$, which typically is of the order of $10^{-1}$, $10^{-2}$, and so on. Subsequently, we randomly select a state variable, referred to as $x$, along with its perturbed counterpart, denoted as $x^{\prime}$. The difference between their values at each iteration, denoted as $\Delta_{t}^i$, is calculated using the expression $|x_{i}-x^{\prime}_{i}|$, where the index $i$ signifies the number of iterations of the state variable. It is essential to ensure that the initial values of $x$ and $x^{\prime}$ are chosen randomly within the $0$ and $1$ range. For instance, suppose we choose $\Delta_{t}^0 = 10^{-3}$ and $x_0 = 0.2$, then $x_0^{\prime}$ would be $0.201$. If $\Delta_{t}$ increases over iterations or stabilizes to a finite value other than zero value (since it cannot grow indefinitely because of the bound behavior of the system), we infer the system has entered a chaotic regime.

\begin{table}[tbh]
\caption{\textcolor{red}{Tabular Representation of TM Method}}
\label{tm}       
\begin{tabular}{lllllllll}
\hline\noalign{\smallskip}
$a$ & $a_1$ & $a_2$ & $a_3$ & $a_4$ & $a_5$ & $a_6$ & $a_6$ & $a_8$  \\
 \noalign{\smallskip}\hline\noalign{\smallskip}
$x$ & $x_1$ & $x_2$ & $x_3$ & $x_4$ & $x_5$ & $x_6$ & $x_7$ & $x_8$ \\
$\Delta_{t}$ & $\Delta_{t}^1$ & $\Delta_{t}^2$ & $\Delta_{t}^3$ & $\Delta_{t}^4$ & $\Delta_{t}^5$ & $\Delta_{t}^6$ & $\Delta_{t}^7$ & $\Delta_{t}^8$ \\
$x^{\prime}$ & $x^{\prime}_1$ & $x^{\prime}_2$ & $x^{\prime}_3$ & $x^{\prime}_4$ & $x^{\prime}_5$ & $x^{\prime}_6$ & $x^{\prime}_7$ & $x^{\prime}_8$ \\
\noalign{\smallskip}\hline
\end{tabular}
\end{table}

To explore the effects of stochasticity, in Table~\ref{tm}, we have shown how to calculate $\Delta_{t}$ up to eight iterations, i.e., $i = 8$). We first have chosen an arbitary initial values, $x$ in the range $(0,1)$ and $\Delta_{t}$, which makes $x^{\rm \prime} = x + \Delta{t}$. We then have evolved $x$ and $x^{\prime}$ independently using the eight random parameter values of $a$ obtained from a distribution. At each iteration, we calculate the $\Delta_{t}^{i}=|x_i-x^{\prime}_i|$.

To calculate the ensemble average of $\Delta_{t}$, we would do the same process mentioned above but with another random initial $x$ value within the range $(0,1)$ and $\Delta_{\rm t}$ will be the same initially. The term $\Delta_{\rm t}^i$ will be calculated for every iteration using the same set of parameters, $a$. For $i$-th iteration, $\Delta_{\rm t}$ will be computed, and the average will be taken with the previous $\Delta_{t}$ at that instant. This process will be continued for many random $x$. In our paper, $10000$ different initial values of $x$ were considered, keeping $\Delta_{\rm t}$ fixed initially. After taking the average of $\Delta_{\rm t}$, these averages ($\Bar{\Delta}_{t}^i$) can be plotted against the iteration number to check the evolution of $\Delta_{\rm t}$.

\subsection{NVN method}
\label{nvn}
\begin{table}[tbh]
\caption{\textcolor{red}{Tabular Representation of NVN Method}}
\label{nvn}       
\begin{tabular}{llllllll}
\hline\noalign{\smallskip}
$a$ & $a_1$ & $a_2$ & $a_3$ & $a_4$ & $a_5$ & $a_6$ & $a_7$  \\
$x$ & $x_1$ & $x_2$  & $x_3$ & $x_4$ & $x_5$ & $x_6$ & $x_7$ \\
$a^{\prime}$ & $a^{\prime}_1$ & $a^{\prime}_2$  & $a^{\prime}_3$ & $a^{\prime}_4$ & $a^{\prime}_5$ & $a^{\prime}_6$ & $a^{\prime}_7$ \\
$x$ & $x^{\prime}_1$ & $x^{\prime}_2$  & $x^{\prime}_3$ & $x^{\prime}_4$ & $x^{\prime}_5$ & $x^{\prime}_6$ & $x^{\prime}_7$ \\
$\Delta_{t}$ & $\Delta_{t}^1$ & $\Delta_{t}^2$ & $\Delta_{t}^3$ & $\Delta_{t}^4$ & $\Delta_{t}^5$ & $\Delta_{t}^6$ & $\Delta_{t}^7$ \\
\noalign{\smallskip}\hline
\end{tabular}
\end{table}

This approach avoids choosing a fixed value for $\Delta_{t}$ at the outset; instead, we initiate the process by randomly selecting an initial value for the state variable $x$ from the range $(0,1)$. Subsequently, we iterate the system using a set of randomly chosen bifurcation parameter values ${a_i}$ from a specific distribution (in our work, we have chosen i = $10000$). To calculate $\Delta_{t}$ at each iteration, we begin with the same initial value $x$ but evolve it with another set of randomly selected bifurcation parameters denoted as ${a^{\prime}_i}$, drawn from the same distribution. Here also, $i$ signifies the number of iterations within different sets. From the above table, we have shown the process for $i = 7$.

For instance, let us consider the case where $x = 0.1$ is the randomly chosen initial value, and the parameter $a$ is selected from a uniform distribution in the range $[0,4]$, resulting in the set $[0.1, 1.9, 2.8, ...]$. We then evolve the system using the same initial value $x = 0.1$ but with another set of bifurcation parameters, such as $[0.3, 2.7, 3.8, ...]$, randomly chosen from the same uniform distribution. For each iteration, we calculate $\Delta_{t}^i=|x_i-x^{\prime}_i|$, where, $\Delta_{t}^1 = 0$.

To compute the ensemble average, we repeat the process as mentioned earlier with another random initial value for $x$, uniformly chosen from the interval $(0,1)$, and iterate the system using different sets of randomly selected bifurcation parameter values obtained from the same distribution. We then calculate $\Delta_{t}^i$ for each iteration and take the average with the corresponding values obtained from the initial set of bifurcation parameters. This process is repeated in our work for $10000$ random initial values of $x$. After computing the average of $\Delta_{t}$ for each iteration, we plot these ensemble averages ($\langle{\Delta_{t}^i}\rangle$) against the number of iterations to analyze the convergence behavior of $\Delta_{t}^i$.

\section{Ergodicity and Convergence}
\label{ergodicity and convergence}
According to Boltzmann's ergodic hypothesis \cite{moore2015ergodic}, given any arbitrary initial condition, the state variable under consideration traverses through all accessible points in phase space over an extended period. In the discrete map represented by Equation~(\ref{eq_2}), the state variable exhibits oscillations within the range of $0.7$ to $2.0$ as shown in the $y$-axis of the Fig.~\ref{map_bif}(a)). The state variable $v_C$ has the minimum and maximum values $0.7$ and $2.0$, respectively. We are considering the numerically obtained bifurcation diagram here because we want to measure the minimum and maximum values accurately. However, the evolution of the state variable of the nonrandom map is not ergodic due to the presence of periodic and chaotic attractors under the parameter variation, eventually causing it to converge to equilibrium points. In order to investigate the ergodic nature of this stochastic state variable, contingent upon distributions with finite endpoints, in this section, we present the ergodicity of the state variable concerning the stochasticity in parameter space in comparison to the nonrandom map described by Equation~(\ref{1D_eq}).

\begin{figure}[tbh]
\centering
\begin{subfigure}[b]{0.5\linewidth}
\includegraphics[width=\linewidth]{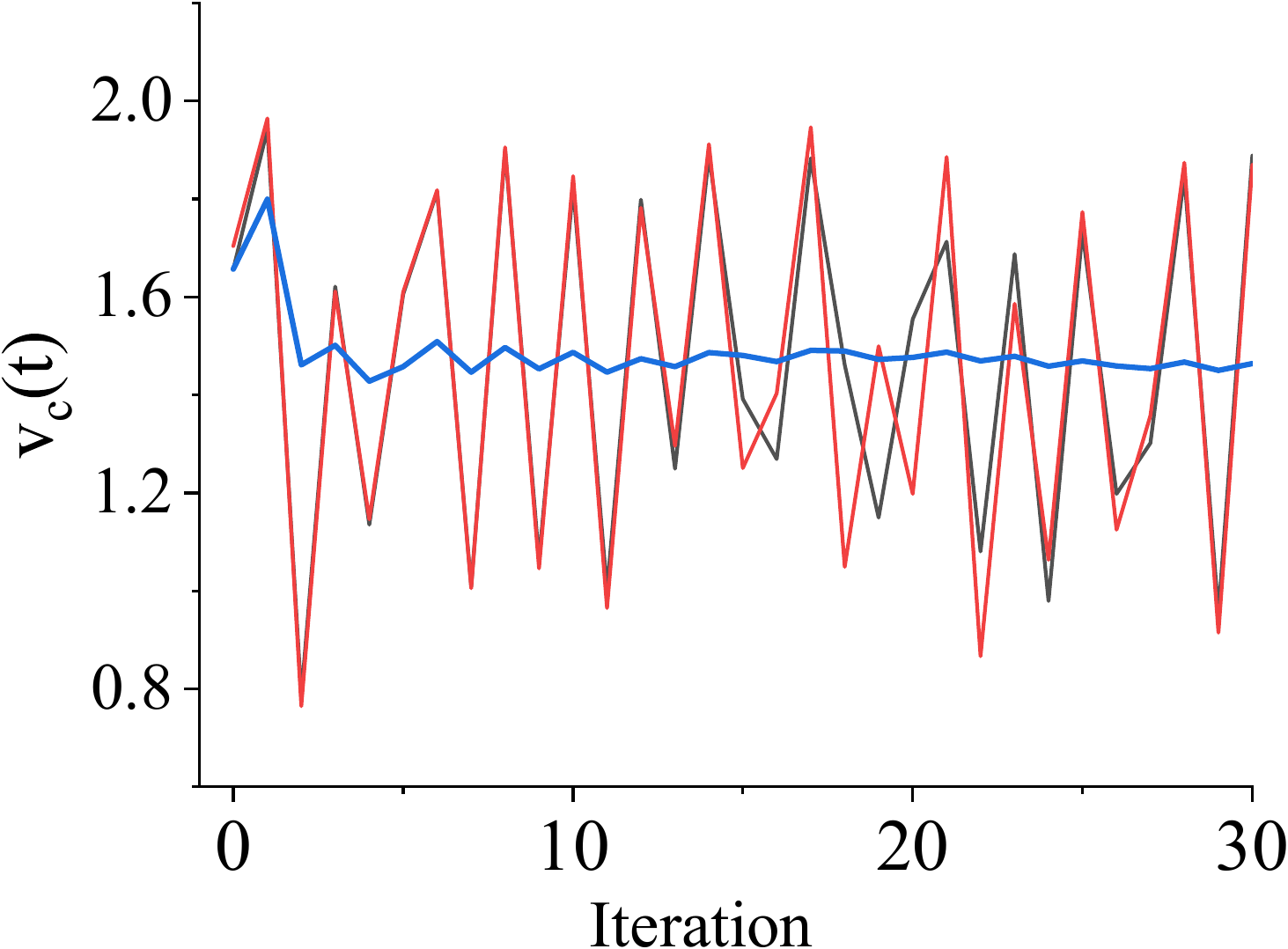}
\caption{}
\end{subfigure}
\begin{subfigure}[b]{0.49\linewidth}
\includegraphics[width=\linewidth]{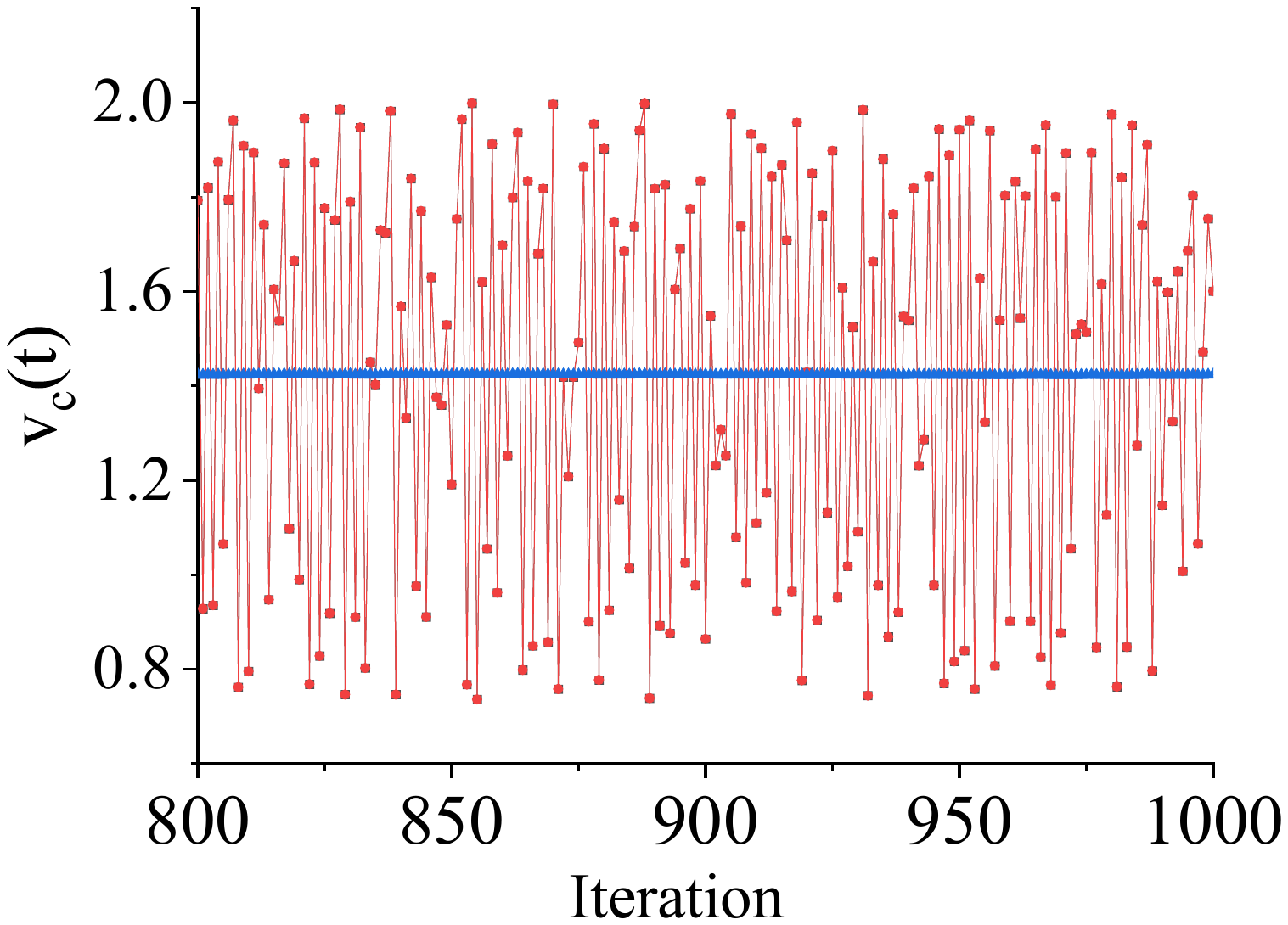}
\caption{}
\end{subfigure} \\
\begin{subfigure}[b]{0.5\linewidth}
\includegraphics[width=\linewidth]{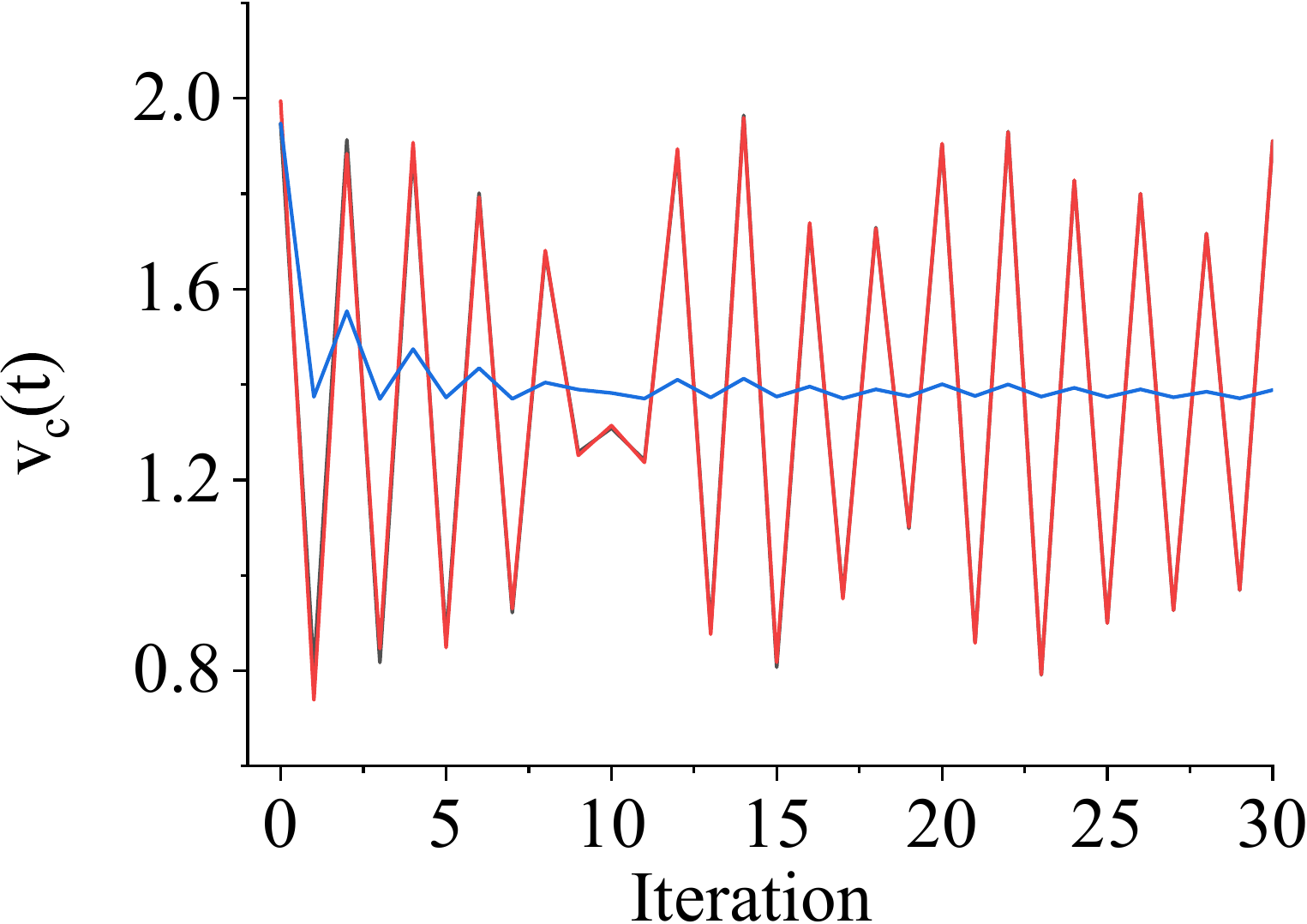}
\caption{}
\end{subfigure}
\begin{subfigure}[b]{0.49\linewidth}
\includegraphics[width=\linewidth]{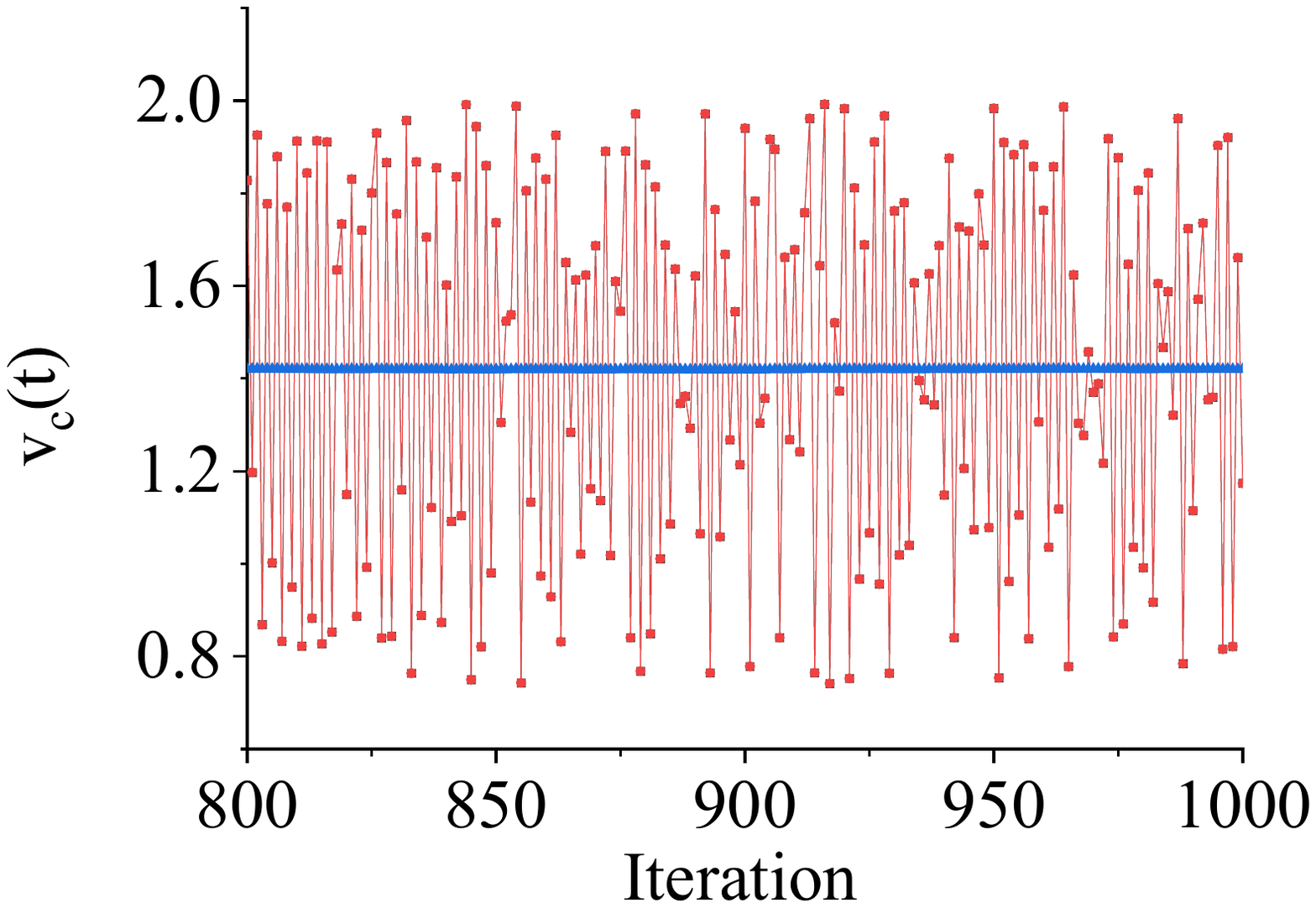}
\caption{}
\end{subfigure}
\caption{TM method results: Two different evolutions of $v_{\rm C}(t)$ with initial separation, $\Delta_{t} = 0.1$ In case of (a) uniform distribution and (c) for symmetric triangular distribution with a range between $2.0$ to $3.5$. (b) and (d) show that after certain iteratons, $\overline{v_{\rm C}(t)}$ reaches to steady state values and $\Delta_t$ goes to $0$. The red and the black colors show the two evolutions of $v_{\rm C}(t)$. The blue color corresponds to the time average of one evolution of $v_{\rm C}(t)$, defined as $\overline{v_{\rm C}(t)}$. (Color Online.)}
\label{FIG:2}
\end{figure}
The nonrandom map exhibits a reverse period incrementing cascade phenomenon with intermittent chaotic windows between periodic attractors while varying the bifurcation parameter $V_1$ in the forward direction, which implies that $V_1$ decreases smoothly, $v_C(t)$ displays an interplay between periodic and chaotic windows. The map possesses the border values in every parameter configuration, where the next iteration will land on the switching surface at $2.0$. The red line in Fig.~\ref{map_bif}(a) is the border under the variation of the parameter $V_1$.

In contrast, in the case of the random map, where the randomness is introduced in the parameter space having the initial value of the state variable $v_C(t)$ randomly chosen between [$0$, $1$], $v_C(t)$ exhibits an ergodic behavior, implying that it does not converge to a fixed periodic attractor (as illustrated in Fig.~\ref{FIG:2}). Consequently, it becomes impossible to predict the $n$-th iteration value of the state variable of the map, $V_n$, before the $n$-th iteration. This stands in stark contrast to the behavior exhibited by the deterministic map, wherein the $n$-th iteration can be readily computed from the piecewise smooth functional form of the map.

An intriguing observation is that when two variables, $v_{\rm C1}$ and $v_{\rm C2}$, are selected, and each is evolved with different bifurcation parameters chosen from a particular distribution at each iteration, they may eventually converge to each other after a certain number of iterations. In other words, defining $|v_{\rm C1} - v_{\rm C2}| = \Delta_{\rm t}$, after the $n$-th iteration (theoretically, as $n\rightarrow\infty$), $\Delta_{t}$ approaches $0$. This phenomenon shows the convergence behavior of the two variables despite the random variations introduced through different bifurcation parameters.

Next, the time average of $v_{\rm C}(t)$, denoted as $\overline{v_{\rm C}(t)}$, is calculated considering the number of iterations. Fig.~\ref{FIG:2}(b) and Fig.~\ref{FIG:2}(d) illustrate that $\overline{v_{\rm C}(t)}$ converges to a fixed value in both cases: when the bifurcation parameter $V_1$ is drawn from a Uniform distribution ($\overline{v_{\rm C}(t)} = 1.423$) and when it is drawn from a Symmetric triangular distribution ($\overline{v_{\rm C}(t)} = 1.415$).

Since $v_{\rm C}(t)$ exhibits an ergodic behavior, we can assume that the state variable possesses $v_{\rm Cmax}$ and $v_{\rm Cmin}$ values in the discrete-time series waveforms. We denote the difference between these extremal values as $\Delta_{v} = (v_{\rm Cmax} - v_{\rm Cmin})$ in the parameter space. Our goal is to examine how $\Delta_{v}$ varies concerning the change in the range of the distribution of bifurcation parameters. This analysis thus determines whether this stochastic map effectively covers all regions in the phase space for a given choice of distribution.

\begin{figure}[tbh]
\centering
\begin{subfigure}[b]{0.5\linewidth}
\includegraphics[width=\linewidth]{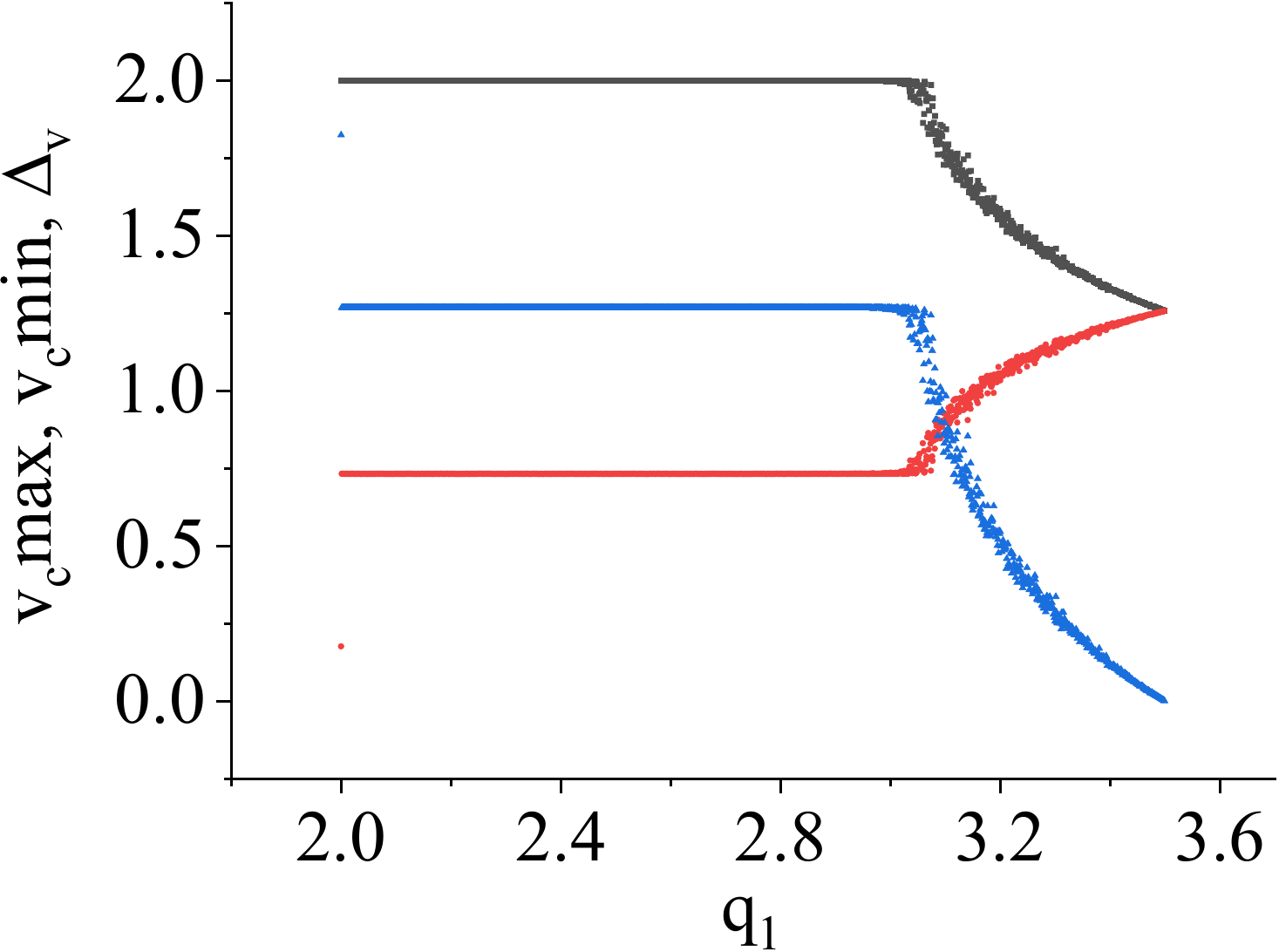}
\caption{}
\end{subfigure}
\begin{subfigure}[b]{0.49\linewidth}
\includegraphics[width=\linewidth]{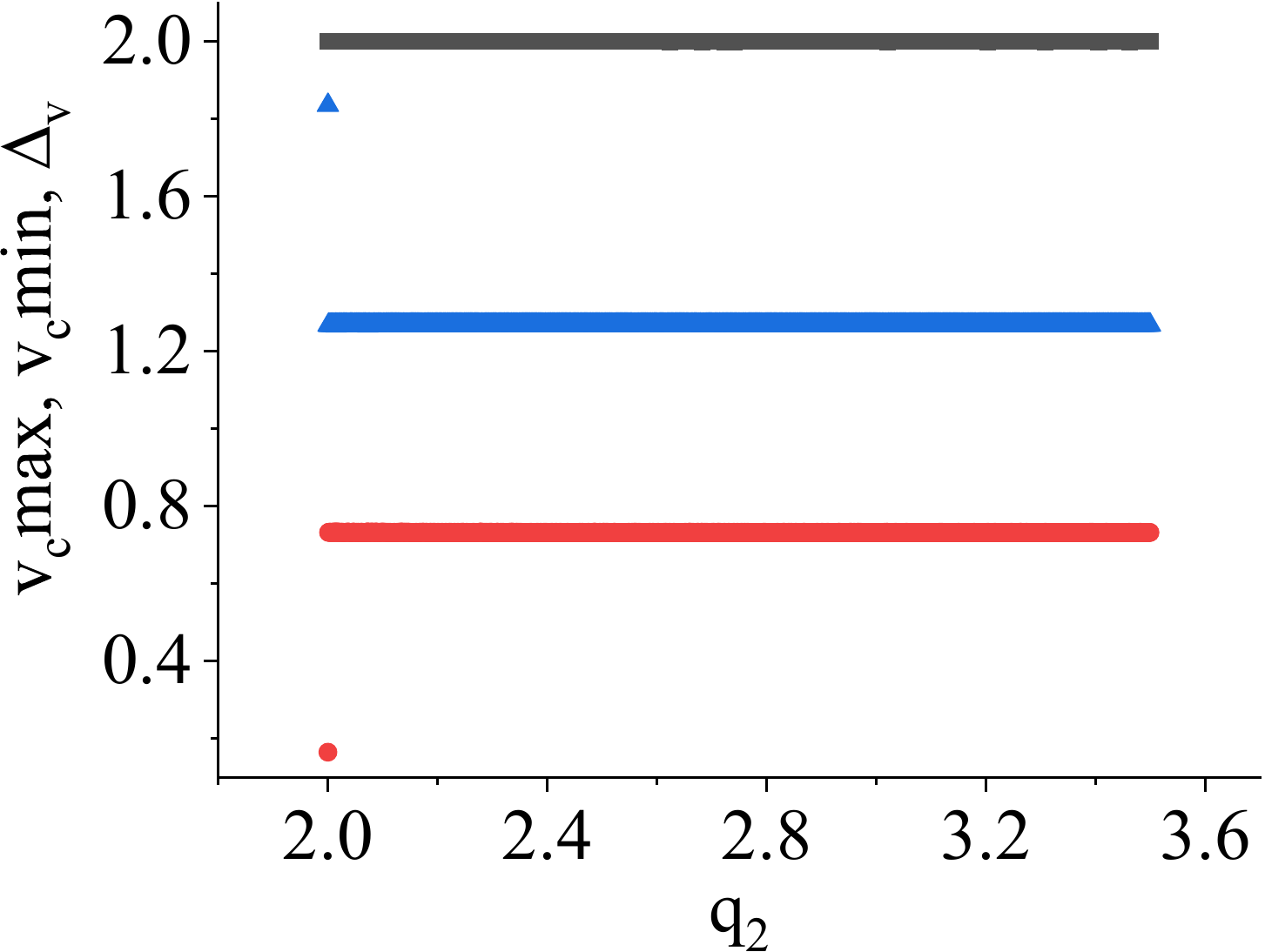}
\caption{}
\end{subfigure}
\begin{subfigure}[b]{0.5\linewidth}
\includegraphics[width=\linewidth]{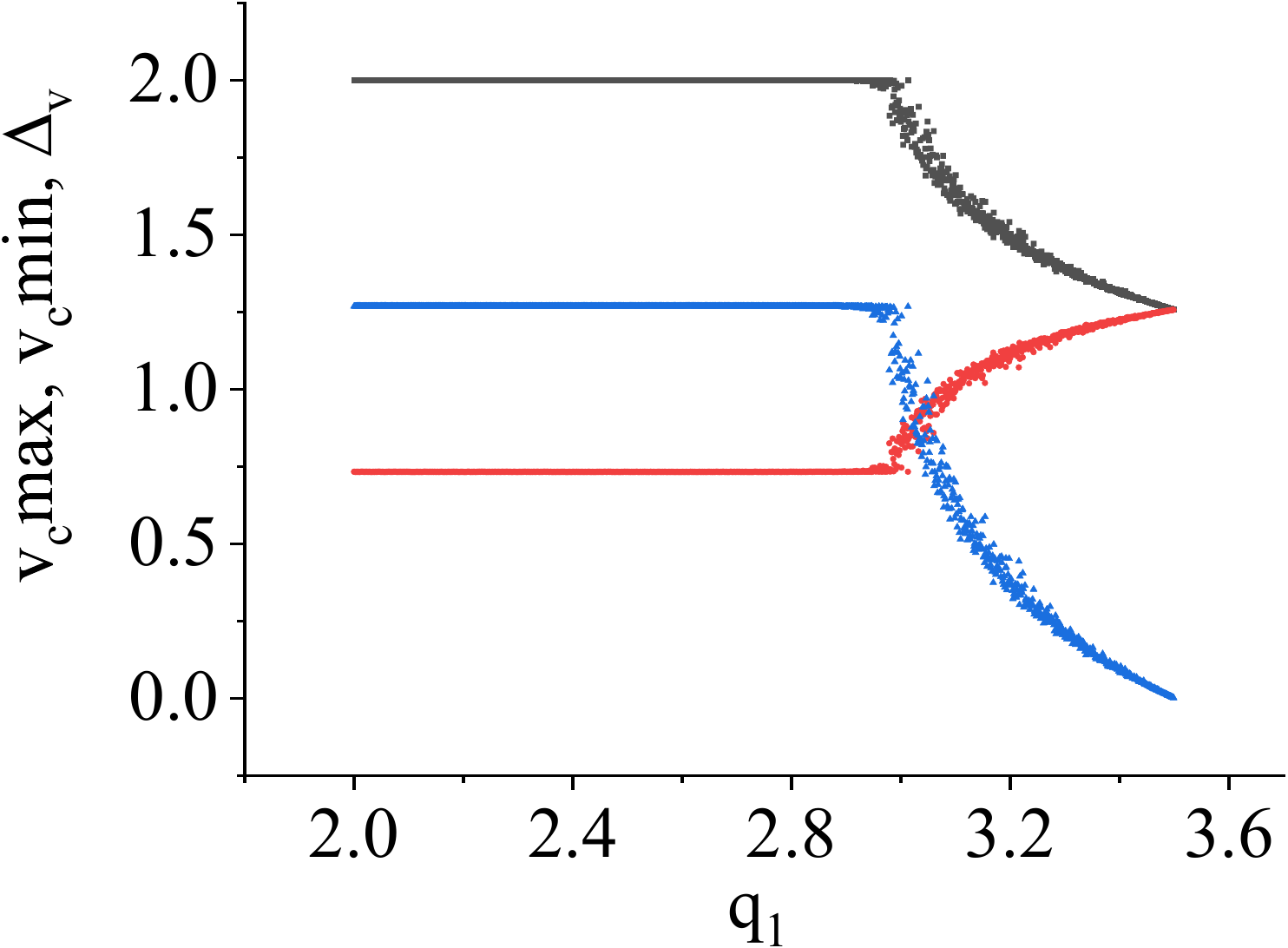}
\caption{}
\end{subfigure}
\begin{subfigure}[b]{0.49\linewidth}
\includegraphics[width=\linewidth]{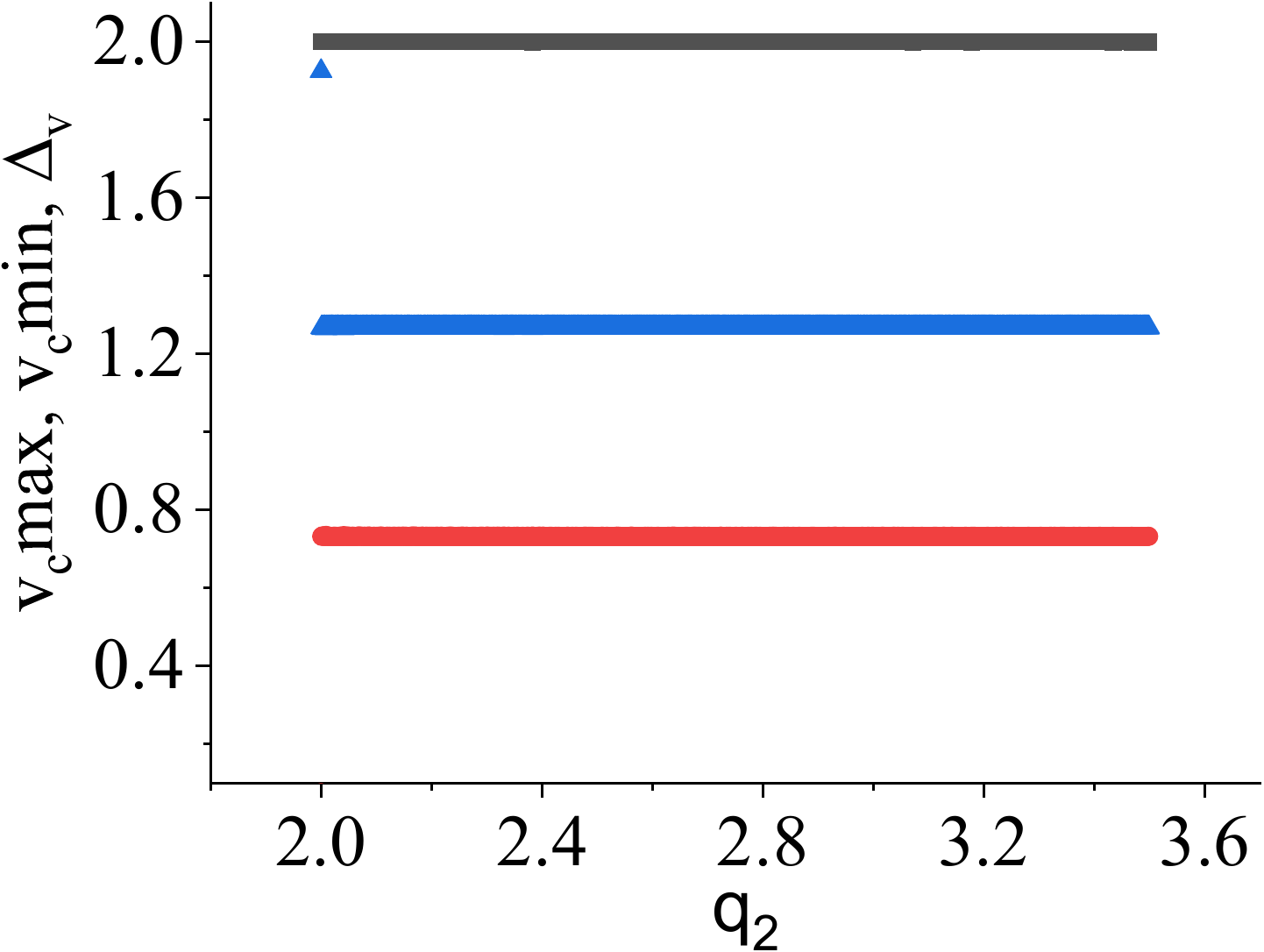}
\caption{}
\end{subfigure}
\caption{TM method results: (a) and (b) show the variations of $v_{\rm Cmax}$, $v_{\rm Cmin}$ and $\Delta_{v}$ for uniform distribution and (c) and (d) show the variations of $v_{\rm Cmax}$, $v_{\rm Cmin}$ and $\Delta_{v}$ for the symmetric triangular distributions. In the case of (a) and (c), $q_1$ is varied from $2.0$ to $3.5$ while $q_2$ is fixed at $3.5$, and in the case of (b) and (d), $q_2$ is varied from $2.0$ to $3.5$ while $q_1$ is fixed at $2.0$. (Color online.)}
\label{FIG:3}
\end{figure}
A uniform distribution between the range $[q_1, q_2]$ for the bifurcation parameter $V_1$ is considered, where $q_1$ and $q_2$ denote the minimum and maximum values of the distribution range, respectively. Subsequently $q_2$ is fixed at $3.5$ while $q_1$ is varied in the range $[2.0, 3.5]$. This generated Fig.~\ref{FIG:3}(a) and Fig.~\ref{FIG:3}(c) where ergodicity of the variable $v_{\rm C}(t)$ for uniform and symmetric triangular distributions are observed.

From Fig.~\ref{FIG:3}(a), it is evident that the ergodicity is consistent up to $q_1 = 3.04$, but beyond that point, it breaks, and the variable converges to a value of $1.25$. This value of $q_1$, where $\Delta_{v}$ changes, closely corresponds to the bifurcation parameter value at which a period-$1$ orbit undergoes a border collision bifurcation and transforms into a period-$1$ orbit ($V_1 \approx 3.1~\mathrm{V}$) for the non-random map. The nature of ergodicity for the symmetric triangular distribution under the same condition is observed next in Fig.~\ref{FIG:3}(c). Notably, the behavior of the ergodicity with the variation of $q_1$ is similar to that of the uniform distribution. Here, the ergodicity breaks at around $q_1 \approx 3.0$.

Finally, the ergodic behavior by fixing $q_1$ at $2.0$ and varying $q_2$ for the uniform distribution and symmetric triangular distribution, as depicted in Fig.~\ref{FIG:3}(b) and Fig.~\ref{FIG:3}(d), respectively, are investigated. In both cases, the ergodicity persists throughout the parameter space with the variation of $q_2$, while $q_1$ remains fixed at $2.0$. This observation implies that there exists a finite likelihood of encountering the state variable $V_n$ within the entirety of the phase space of the map, where the distribution spans the interval between $q_1$ and $q_2$.
 
\section{Invariant Probability Density Function}
\label{invariant probability}
In \cite{PhysRevE.96.042139}, it was demonstrated that the distribution of the state variable for a stochastic map converges to a Maxwell-Boltzmann distribution in the limit of a long time. In contrast, the distribution observed in our specific stochastic map closely resembles the shape of the invariant measure associated with a logistic map \cite{layek2015introduction}.

\begin{figure}[tbh]
\centering
\begin{subfigure}[b]{0.7\linewidth}
\includegraphics[width=\linewidth]{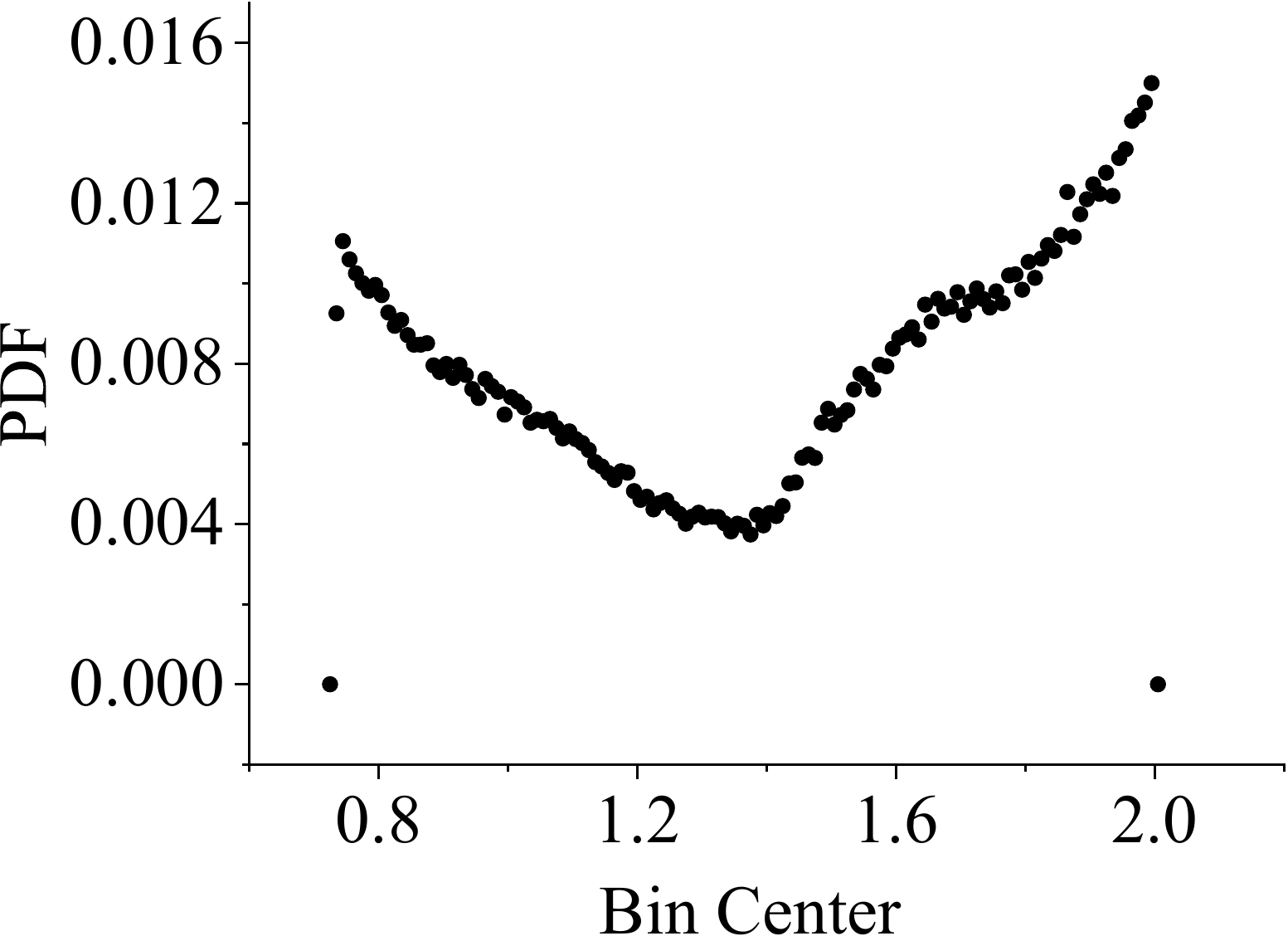}
\caption{}
\end{subfigure}
\begin{subfigure}[b]{0.7\linewidth}
\includegraphics[width=\linewidth]{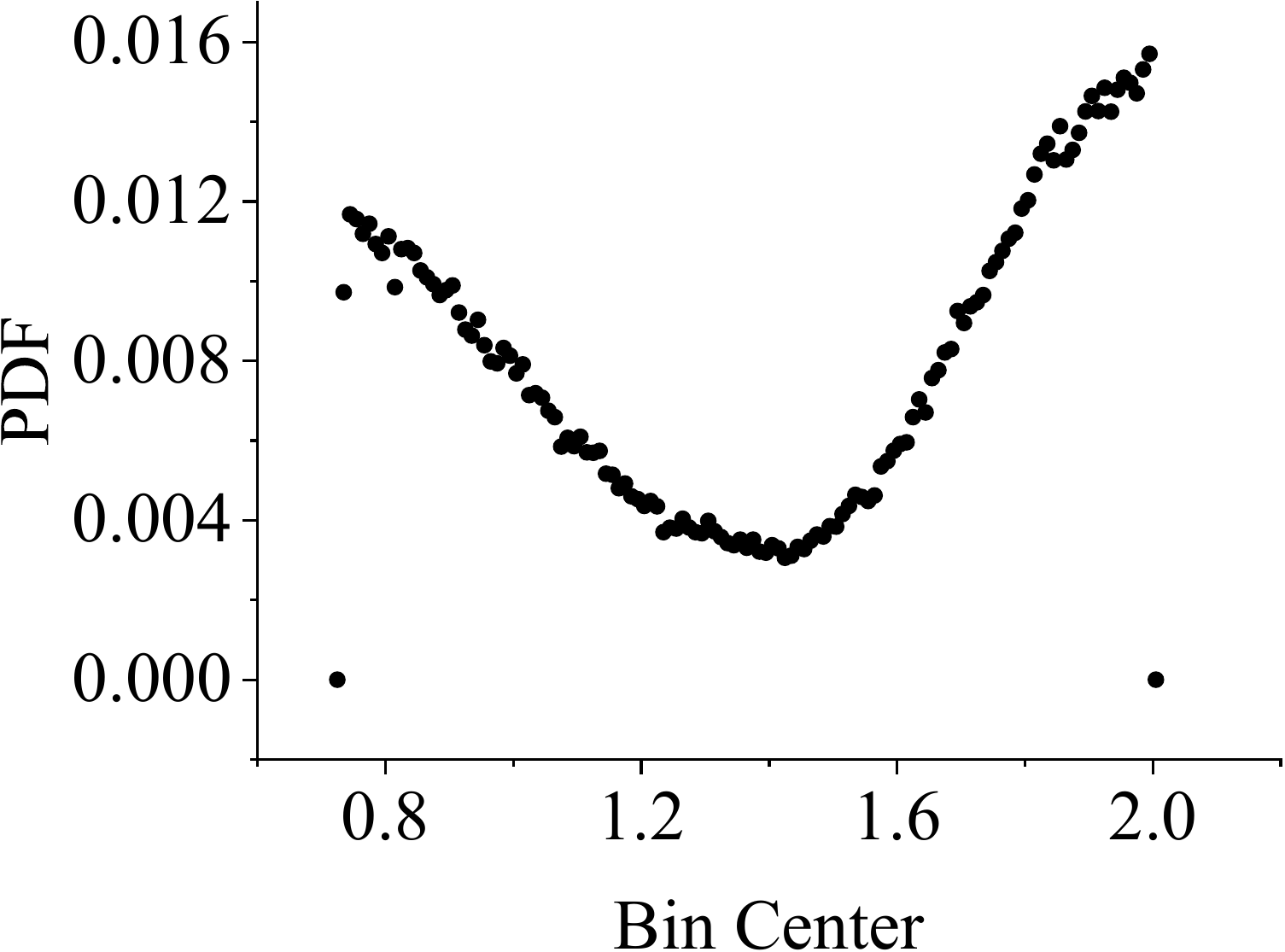}
\caption{}
\end{subfigure}
\caption{The probability density functions in the parameter space (a) for uniform distribution and (b) for symmetric triangular distribution. The bifurcation parameter is considered within [$2.0$,$3.5$] for both distributions.}
\label{ipd}
\end{figure}
Using uniform and symmetric triangular probability distributions, the probability density function (PDF) of the state variable of the map is plotted, as shown in Fig.~\ref{ipd}(a) and Fig.~\ref{ipd}(b), respectively. Notably, these PDFs remain invariant regardless of the initial conditions when the bifurcation parameter is drawn from a fixed distribution with predefined boundary points, as stated in the caption of Figure~\ref{ipd}.

Fig.~\ref{ipd}(a) and Fig.~\ref{ipd}(b) visually demonstrate the probabilistic behavior of the system for both distributions. As we observe the values along the $x$-axis, we notice that the distribution function progressively decreases until reaching a certain point and then begins to increase. This trend indicates that the random state variable is more likely to be situated in the vicinity of the switching surface at $2.0$ V, and the function is more stable there. The dips in both curves signify that the probability of obtaining the variable $v_{\rm C}$ is relatively lower, around $1.5$, and the function is less stable.

It is observed numerically that as long-run iteration progresses, the distribution of $p_t(v_C)$ (which represents the probability density function of $v_C$ at time $t$) ceases to evolve with time and instead converges towards a fixed distribution denoted as $p(v_C)$. In analogy with the Frobenius-Perron operator used for any deterministic map, we can define a similar operator to derive an invariant measure for this stochastic map.

The expression for the stationary distribution $p(x)$ as $n$ tends to infinity is given by:

\begin{equation}
p(x) \stackrel{n\rightarrow\infty}{=} \int dy \int_{q_1}^{q_2} dV_1  p(y) \delta(x - g(y,V_1))
\label{eq_4}
\end{equation}
Where $x = v_{\rm C}(n+1)$ and $y = v_{\rm C}(n)$. Although $n\rightarrow\infty$, $p(v_{\rm C}(n+1))\approx p(v_{\rm C}(n))$, we can put $p(x)$ in place of $p(y)$ in the above equation, to find the $p(x)$. 

In Equation~(\ref{eq_4}), we can observe that the stationary distribution $p(x)$ is dependent on the parameters $q_1$ and $q_2$, representing the endpoints of the probability distribution, as well as the specific type of probability distribution employed. This dependence has been further corroborated through numerical investigations.

\section{Evolution of stochastic map}
\label{evolution}

Upon scrutinizing the dynamics of this probabilistic map, it has come to light that there exist two distinct regimes of evolution:

\begin{itemize}
    \item Non-Chaotic Regime
    \item Chaotic Regime
\end{itemize}

Given the stochastic nature of this map, it is precluded from predicting the $n$-th iteration value of the state variable $V_n$ prior to the $n$-th iteration. Consequently, the analytic computation of fixed points and their corresponding stability, contingent on this stochastic nature, remains an infeasible task. In light of this, we have undertaken numerical computations for two diverse evolutions of state variables within the probabilistic map. These computations involve selecting different initial conditions while maintaining a consistent set of parameter values derived from the distribution for every iteration (referred to as the TM method) or alternatively, maintaining the same initial condition while selecting a distinct set of parameter values obtained from a distribution (referred to as the NVN method).

Should the difference between the two state variables ($\Delta_t$) approach zero asymptotically, or should the two evolutions converge in later iterations, this scenario is designated as a periodic or non-chaotic regime. It is worth noting that for the prediction of periodicity, we must observe a minimum of two distinct evolutions. Conversely, suppose the discrepancy between the two state variables ($\Delta_t$) does not asymptotically approach zero but instead attains a fixed nonzero value. In that case, this scenario is classified as a chaotic regime, as the two evolutions ultimately fail to converge in subsequent iterations \cite{doi:10.1142/S0129183115500862}.

\section{Results for TM Method}
\label{results_tm}
\subsection{Non-Chaotic regime}
\label{Non-Chaotic regime}
In Fig.~\ref{FIG:2}, we have shown a pair of initial values of $v_C(t)$, which are slightly different and allow them to evolve as a function of time within the parameter [2.0,3.5]. We have found that $v_C$ does not attain a fixed point value. But when the separation, $\Delta_t$, between two state variables gradually decays towards zero after certain iterations, we can say that the system approaches towards the nonrandom behavior. The region in parameter space where this condition occurs is called a regular or non-chaotic regime. When $\Delta_t$ is zero, we can say that there is no damage to the state variable of the random map. \cite{KHALEQUE2014599}.

\begin{figure}[tbh]
\centering
\begin{subfigure}[b]{0.7\linewidth}
\includegraphics[width=\linewidth]{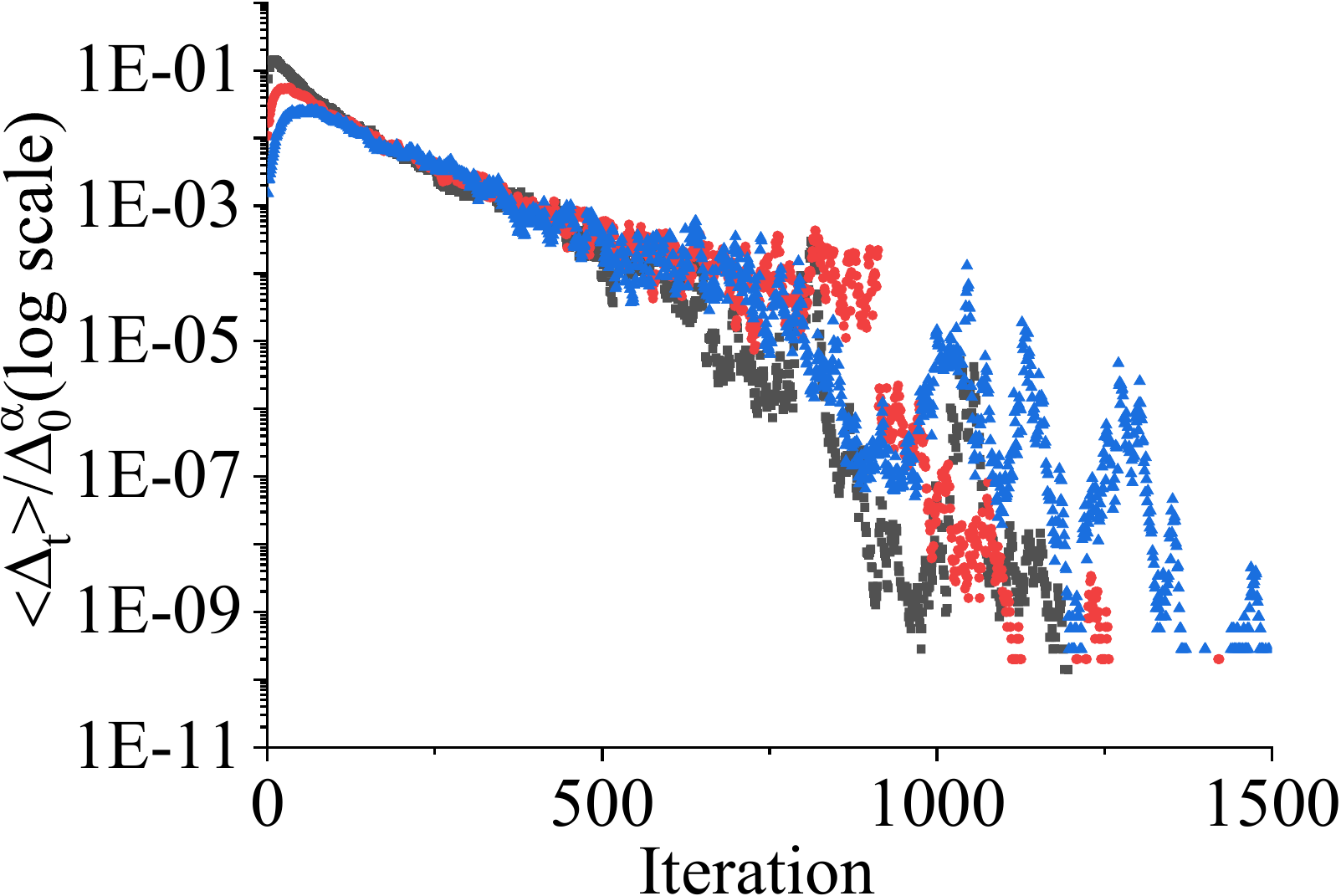}\caption{}
\end{subfigure}
\begin{subfigure}[b]{0.7\linewidth}
\includegraphics[width=\linewidth]{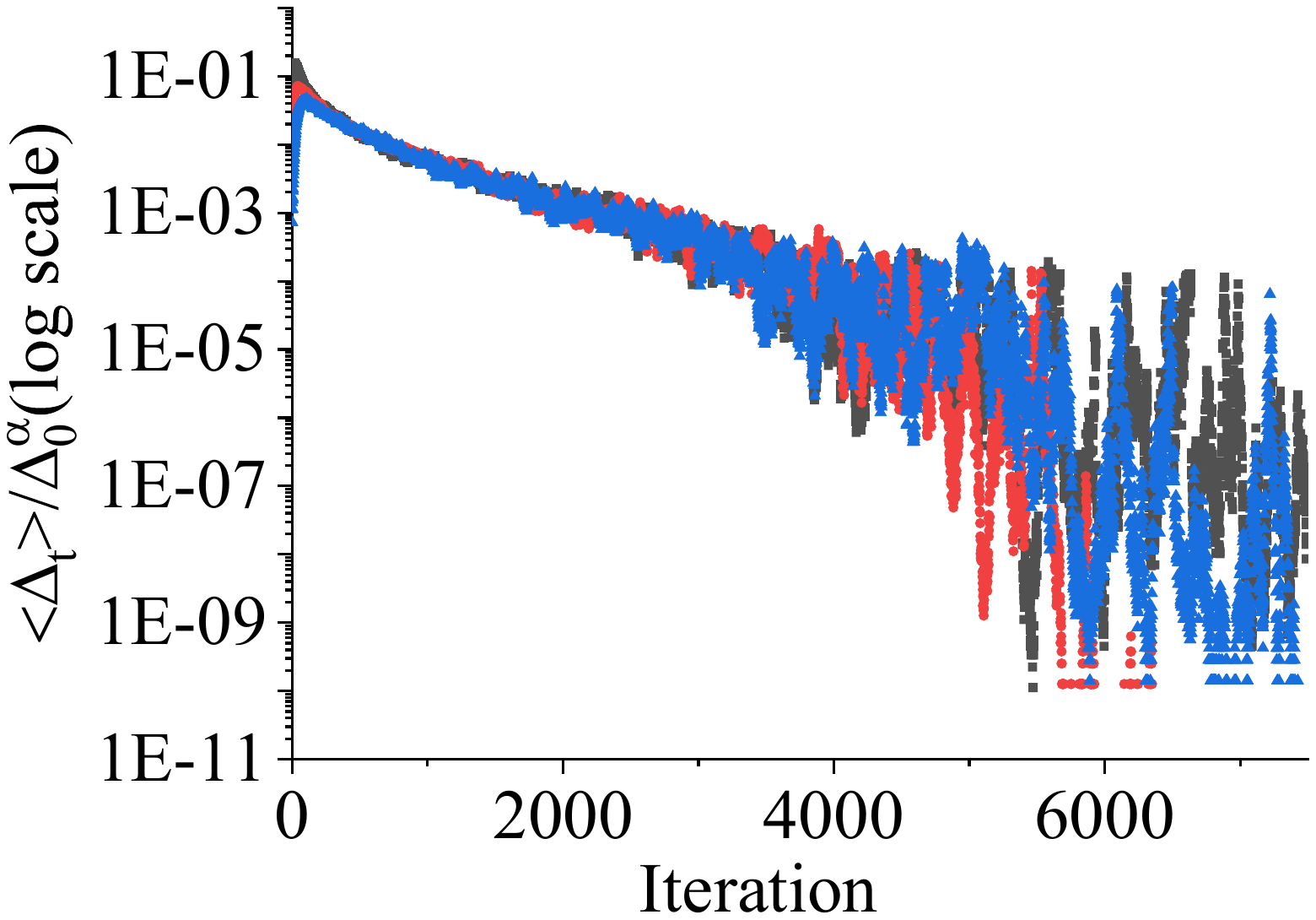}
\caption{}
\end{subfigure}
\caption{TM method results: The scaled value of damage $\Delta_{t}$ against the no of iterations in the case of (a) uniform and (b) symmetric triangular distributions with a range of $2.0$ to $3.5$. The black color is for the initial separation value between two variables, $\Delta_0$ is $0.1$, the red curve is for $\Delta = 0.01$, and the blue curve is for $\Delta=0.001$. (Color online.)}
\label{FIG:5}
\end{figure}
In a non-chaotic regime, when starting from an initial random value of $\Delta_0$ and measuring the difference $\Delta_{\rm t}$ between two independently chosen variables $v_{\rm C1}$ and $v_{\rm C2}$ after a certain number of iterations, it eventually converges to zero. To quantify this behavior, we numerically demonstrate the scaled value of ensemble averages of $\Delta_{\rm t}$, representing the amount of convergence with the number of iterations using the TM method, as shown in Fig.~\ref{FIG:5}. Fig.~\ref{FIG:5}(a) illustrates the variation of damage with the number of iterations for the uniform distribution, and Fig.~\ref{FIG:5}(b) depicts the same for symmetric triangular distribution.

In both cases depicted in Fig.~\ref{FIG:5}, we examine the behavior of the scaled value of the damage for three initial random numbers of $\Delta_{\rm t}$ (the specific values are provided in the caption of Fig.~\ref{FIG:5}). Initially, the three different initial scaled values of $\Delta_t$ increase up to a certain number of iterations, and then they start to decrease as the iterations progress. Eventually, these scaled values of $\Delta_{\rm t}$ merge together, exhibiting a convergence towards a common trajectory. After some iterations, the scaled value of $\Delta_{\rm t}$ demonstrates oscillatory behavior in each figure. Despite this oscillation, it is important to note that $\Delta_{\rm t}$ in each case converges towards zero as time progresses. It shows that although the state variable $v_C(t)$ does not attain a periodic behavior within the parameter range [$2.0$,$3.5$], $\Delta_t$ between the state variable and its slightly different value for three different $\Delta_t$ values decay towards zero as time progresses. It shows the ordered behavior of the random map in the parameter range, as mentioned earlier. Each plot in Fig.~\ref{FIG:5} demonstrates that the time-averaged scaled value of damage, denoted by $\langle\Delta_{\rm t}\rangle$, exhibits monotonically decreasing value towards zero for both distributions, even though they have the same distribution range. This behavior suggests a regular or non-chaotic nature in the system.

Furthermore, the relationship between $\Delta_{\rm t}$ and $\Delta_0$ is nonlinear, specifically given by $\langle\Delta_{\rm t}\rangle \propto {\Delta_0^\alpha} \exp (\lambda t)$, where $\Delta_0$ is the initial separation between two state variables, $v_{\rm C}(t)$. In contrast, for the nonrandom version of this map, it follows $\Delta_{\rm t} \propto \Delta_0$ $\exp(\lambda t)$. The exponent $\alpha$ is estimated to be $0.15\pm0.01$ for the uniform distribution and $0.05\pm0.01$ for the symmetric triangular distribution. The Lyapunov Exponent ($\lambda$) depends strongly on the endpoints, $q_1$, $q_2$, of the distributions. It shows an increase in magnitude as $q_2 - q_1$ is decreased. The corresponding approximated Lyapunov exponents are approximately $\lambda \simeq -5 \times 10^3$ for the uniform distribution and $\lambda \simeq -1.64 \times 10^3$ for the symmetric triangular distribution. The distributions were selected from the range of $2.0$ to $3.5$. The approximated Lyapunov exponents represent the rates of separation of initially close trajectories in phase space. A more negative Lyapunov exponent signifies a faster convergence of nearby trajectories, indicating greater stability in the system.

These findings provide valuable quantitative insights into the relationship between the initial scaled damage $\Delta_0$ and its subsequent evolution over time, characterized by the time-averaged scaled damage $\langle\Delta_{\rm t}\rangle$ for the random considered map. The observed nonlinear relationship with the exponent $\alpha$ indicates the sensitivity of the response of the system to the initial conditions, especially when considering random distributions.

\begin{figure}[tbh]
\centering
\begin{subfigure}[b]{0.7\linewidth}
\includegraphics[width=\linewidth]{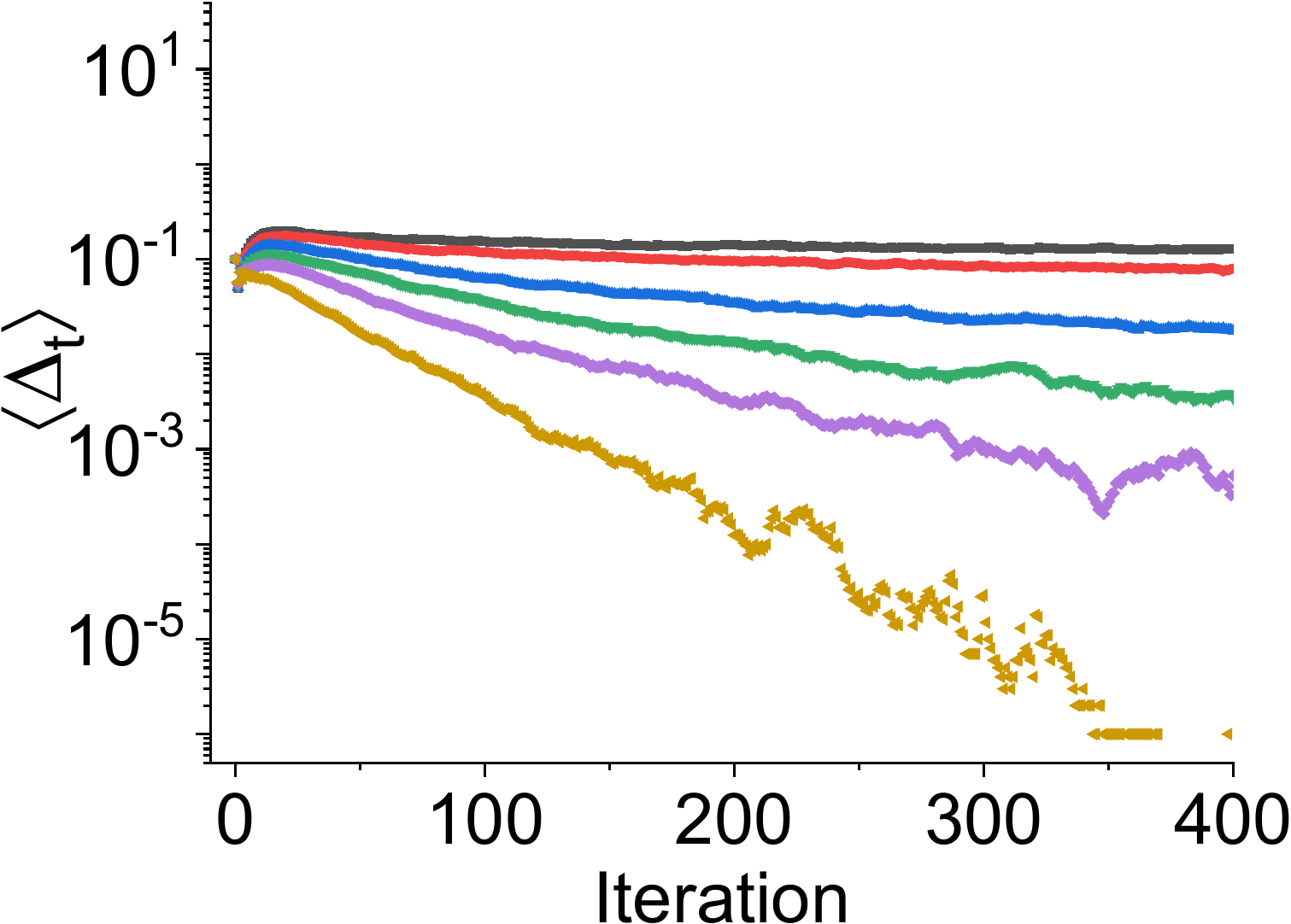}
\caption{}
\end{subfigure}
\begin{subfigure}[b]{0.7\linewidth}
\includegraphics[width=\linewidth]{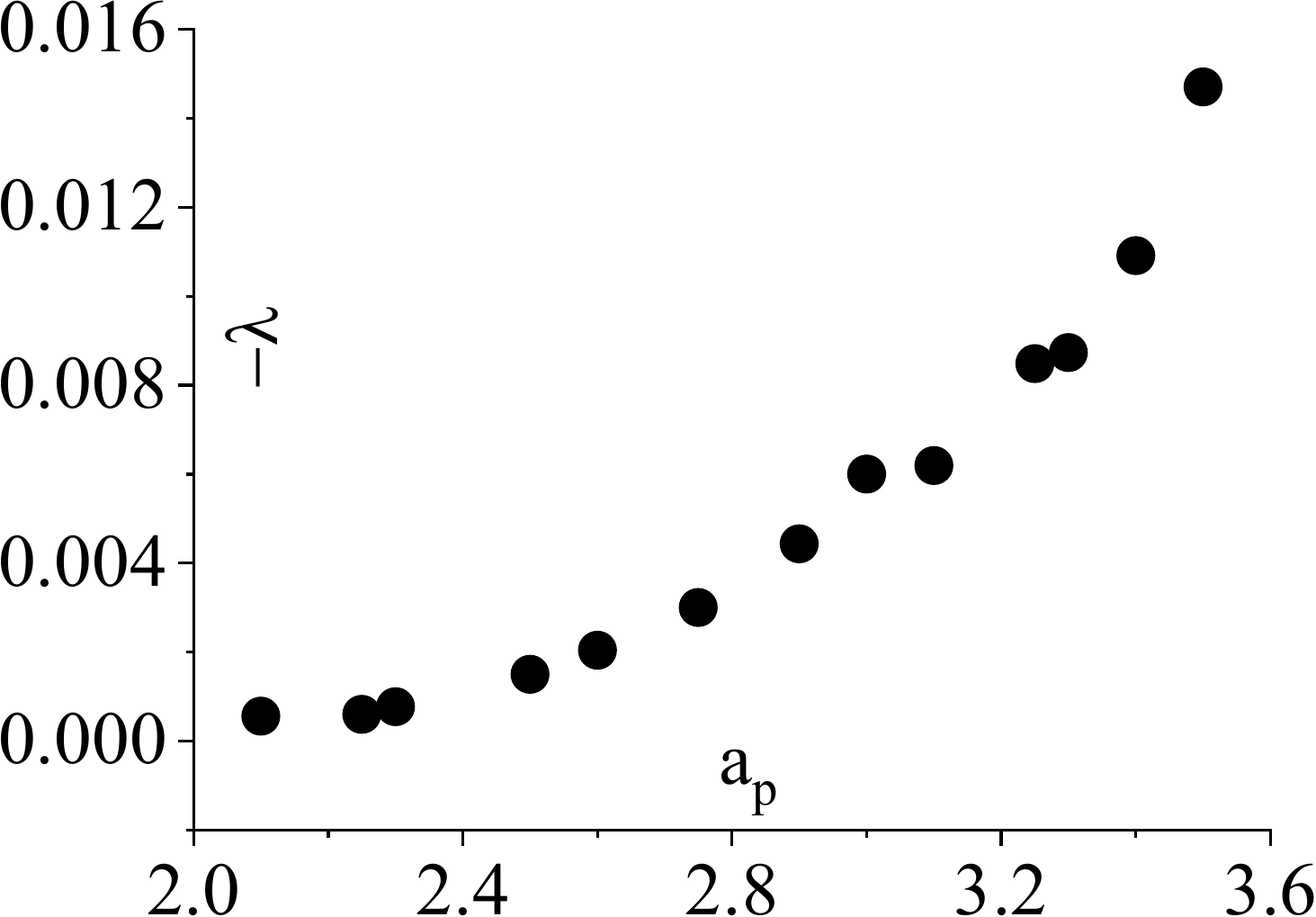}
\caption{}
\end{subfigure}
\caption{TM method results: $\Delta_{t}$ against time for the asymmetric triangular distribution with different peaks at $a_p$. Black color is for $a_p = 2.25$, Red color is for $a_p = 2.5$, Blue color is for $a_p = 2.75$, Green color is for $a_p = 3.0$, Purple color is for $a_p = 3.25$, Yellow color is for $a_p = 3.5$. (b) the variation of the Lyapunov exponent with the values of $a_p$ for the asymmetric triangular distribution peaked at $ a_p $ in the range between $2.0$ to $3.5$ (Color Online.)}
\label{FIG:6}
\end{figure}
In Fig.~\ref{FIG:6}(a), the variation of $\langle\Delta_t\rangle$ with time is presented for asymmetric triangular distributions, where the distribution ranges from $2.0$ to $3.5$. Six different peaks were considered to quantify the asymmetries, specifically $2.25 \leq a_p \leq 3.5$. Notably, it is observed that as $a_p$ increases, the negative slope of $\Delta_t$ with time also increases, indicating a more rapid convergence of $\Delta_t$ towards zero.

Fig.~\ref{FIG:6}(b) shows the relationship between the Lyapunov exponent ($\lambda$) and $a_p$. The plot reveals that as $a_p$ progressively increases, the modulus of $\lambda$ also increases exponentially. The interpretation of $\lambda$ suggests that as the value of $a_p$ decreases, $\Delta_t$ vanishes more gradually. Conversely, as $a_p$ increases, $\Delta_t$ becomes more prominent, signifying a higher degree of sensitivity to initial conditions and potentially less stable behavior.

\subsection{Chaotic regime}
\label{chaotic regime}
In the previous case, we observed that $\langle\Delta_{t}\rangle$ approaches zero with increasing iterations for specific values of $q_1 = 2.0$ and $q_2 = 3.5$. However, when we vary $q_2$ while keeping $q_1$ fixed at $2.0$, certain scenarios arise where $\langle\Delta_{t}\rangle$ does not reach zero. Instead, it asymptotically converges to a constant nonzero value. This persistent behavior of ergodicity throughout the parameter space, with the system settling into an asymptotic non-zero value of $\langle\Delta_{\rm t}\rangle$, is defined as the ``chaotic regime." We denote the asymptotic non-zero value of $\langle\Delta_{\rm t}\rangle$ as the saturation of $\Delta_t$, represented as $\Delta_{\rm sat}$.

For this specific stochastic map, when a pair of variables $v_{\rm C1}$ and $v_{\rm C2}$ are chosen within the chaotic regime, their difference $\Delta_{t}$ exhibits dynamic changes with time. Initially, $\Delta_{t}$ is zero, indicating no ergodicity for a certain period. Consequently, the waveforms $v_{\rm C1}$ and $v_{\rm C2}$ effectively merge into each other during this time interval. However, after this initial phase, the separation $\Delta_t$ assumes a specific nonzero value, causing the two variables to diverge. This separation persists for a certain duration until $\Delta_{t}$ decreases back to zero again, and the waveforms merge once more. This cyclic behavior repeats itself as the iterations progress forward. This behavior pattern suggests that, within the chaotic regime, there exists a dynamic interplay involving the amalgamation and division of two state variables over time. This particular behavior has not been previously observed in the context of a stochastic continuous map and arises solely as a consequence of the non-smooth nature of the map.

\begin{figure}[tbh]
\centering
\begin{subfigure}[b]{0.5\linewidth}
\includegraphics[width=\linewidth]{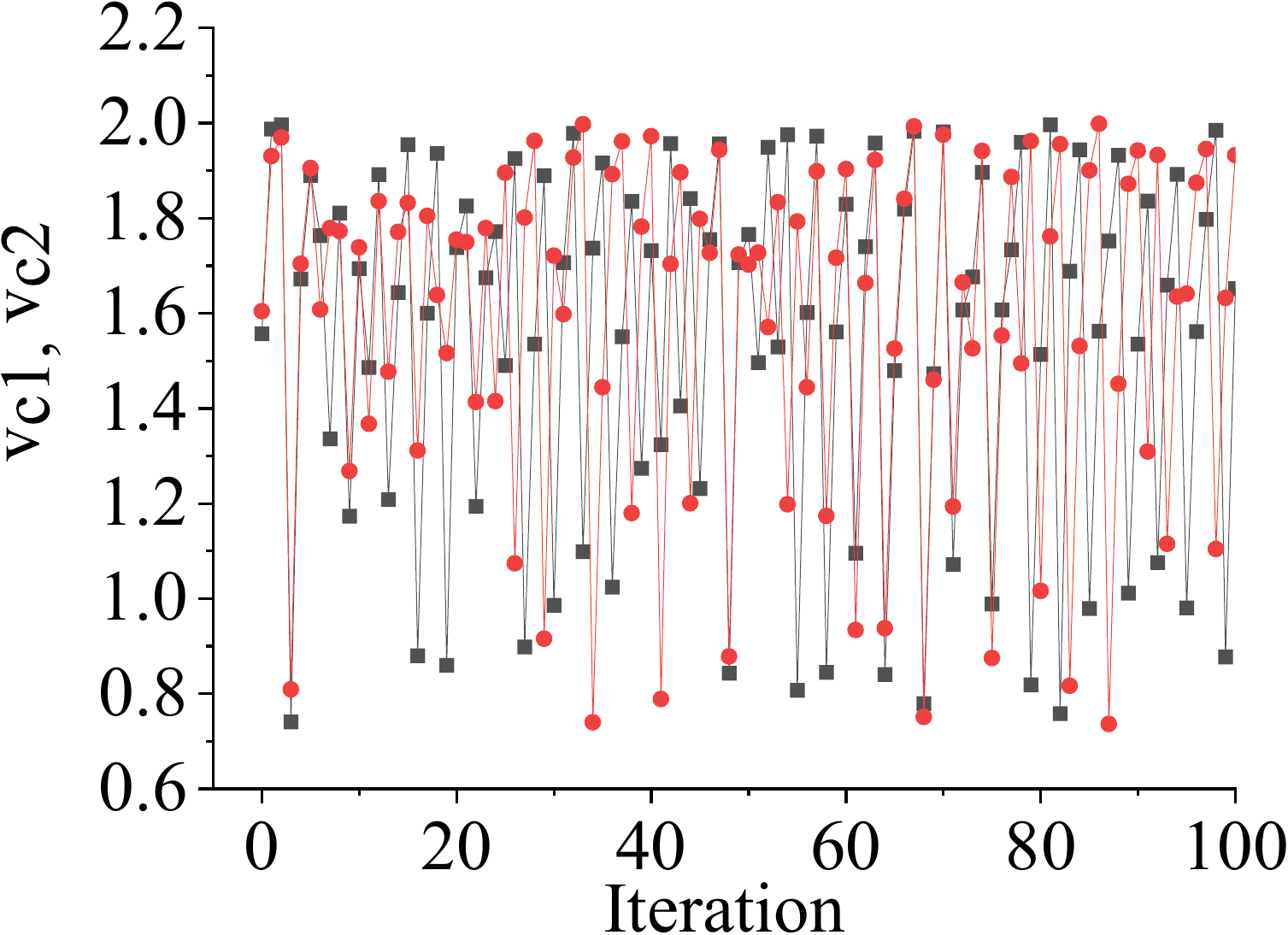}
\caption{}
\end{subfigure}
\begin{subfigure}[b]{0.49\linewidth}
\includegraphics[width=\linewidth]{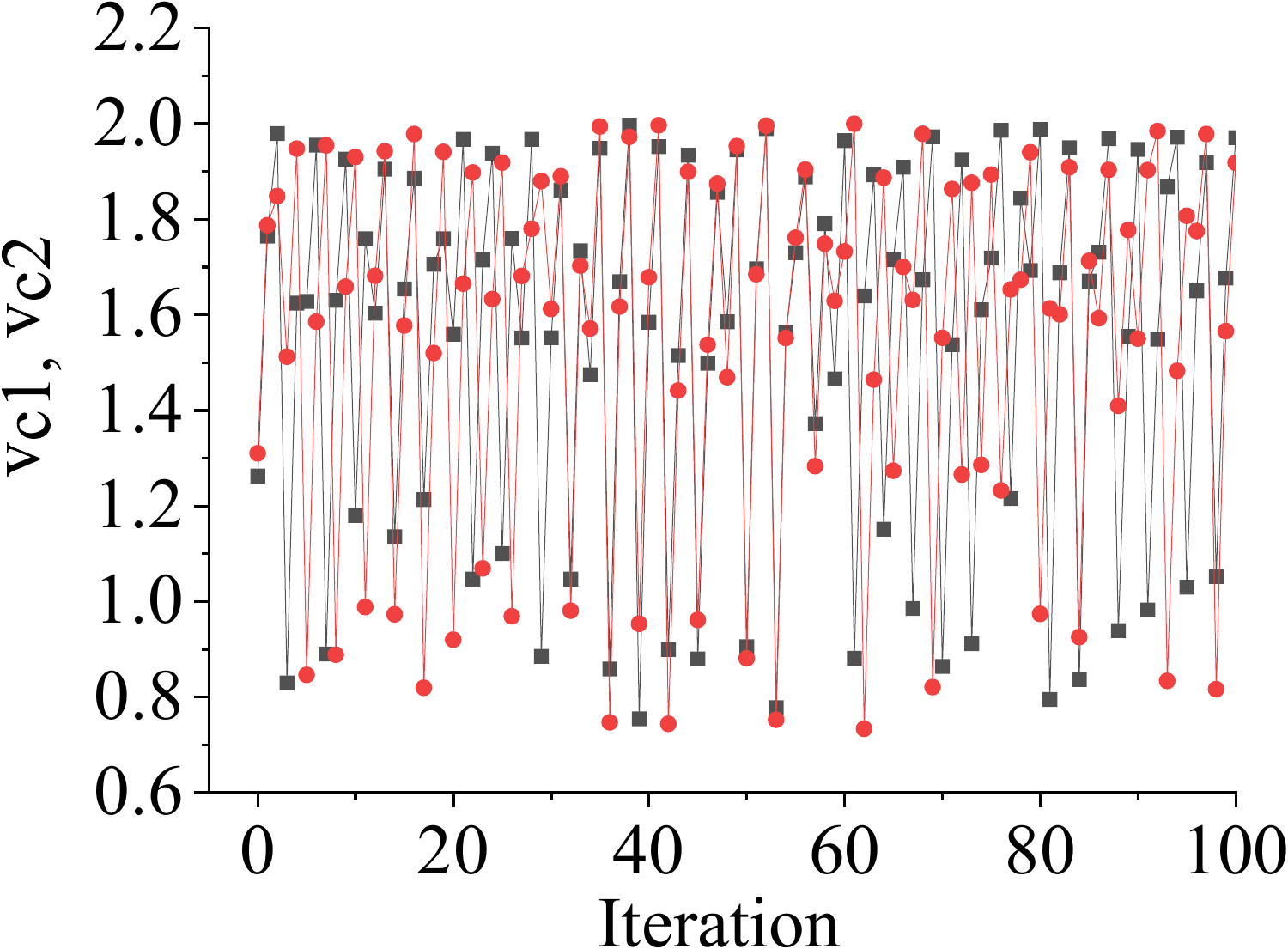}
\caption{}
\end{subfigure} \\
\begin{subfigure}[b]{0.5\linewidth}
\includegraphics[width=\linewidth]{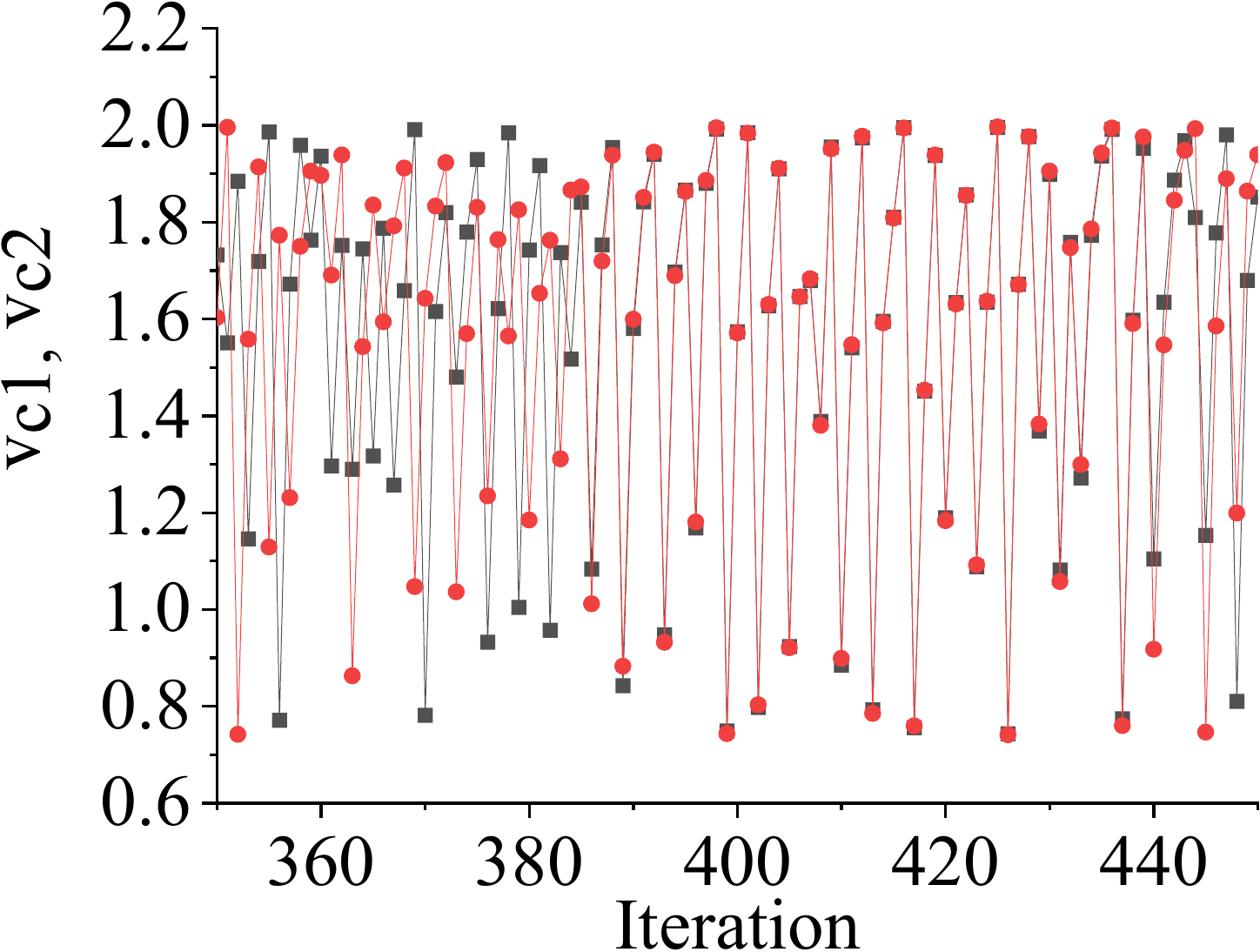}
\caption{}
\end{subfigure}
\begin{subfigure}[b]{0.49\linewidth}
\includegraphics[width=\linewidth]{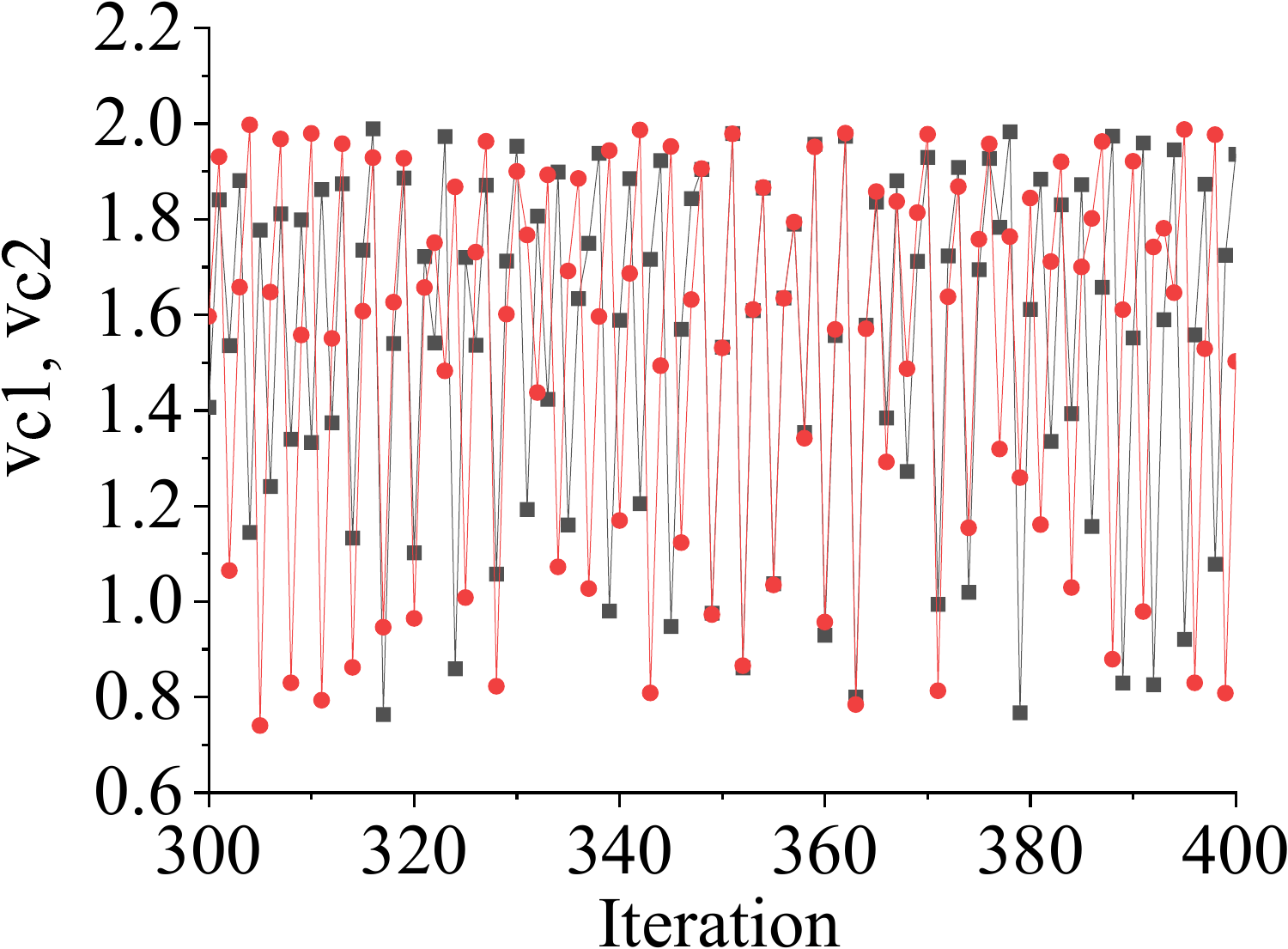}
\caption{}
\end{subfigure}
\caption{TM method results: \textbf{Uniform Distribution}: Two different evolutions of $v_{\rm C}(t)$ when (a) $\Delta_{t}$ is nonzero, (b) when $\Delta_{t}$ is zero. \textbf{Triangular Distribution}: Two different evolutions of $v_{\rm C}(t)$ when (a) $\Delta_{t}$ is nonzero, (b) when $\Delta_{t}$ is zero. Black and red colors are for $v_{\rm C1}$ and $v_{\rm C2}$, respectively. (Color online.)}
\label{FIG:7}
\end{figure}
In Fig.~\ref{FIG:7}, we illustrate the alternating nonzero and zero values of $\Delta_t$ for both uniform and symmetric triangular distributions. The time-series waveforms of the two-state variables, denoted by red and black colors, erratically merge and separate as time progresses. This pattern emerges due to the non-smoothness property of the stochastic map, where there are periodic windows having finite width in between chaotic attractors in the bifurcation diagram in the case of the nonrandom map.

It is worth noting that in the case of the nonrandom map, which exhibits a reverse period incrementing cascade phenomenon with chaotic windows between periodic orbits (as depicted in Fig.~\ref{map_bif}(b)), the ratios of the widths of successive periodic windows and subsequent chaotic windows converge to a constant value \cite{iiserkeprints1110}. However, in contrast, for the stochastic map, the ratios of consecutive spans where $\Delta_{t}$ is zero and the ratios of successive spans with nonzero $\Delta_t$ are random and do not converge to a fixed value, as observed in the nonrandom map.

The above-mentioned findings emphasize the distinct and complex nature of the behavior of the random map in the chaotic regime, showcasing the significance of randomness in shaping its dynamics and ergodic properties.

\begin{figure}[tbh]
\centering
\begin{subfigure}[b]{0.7\linewidth}
\includegraphics[width=\linewidth]{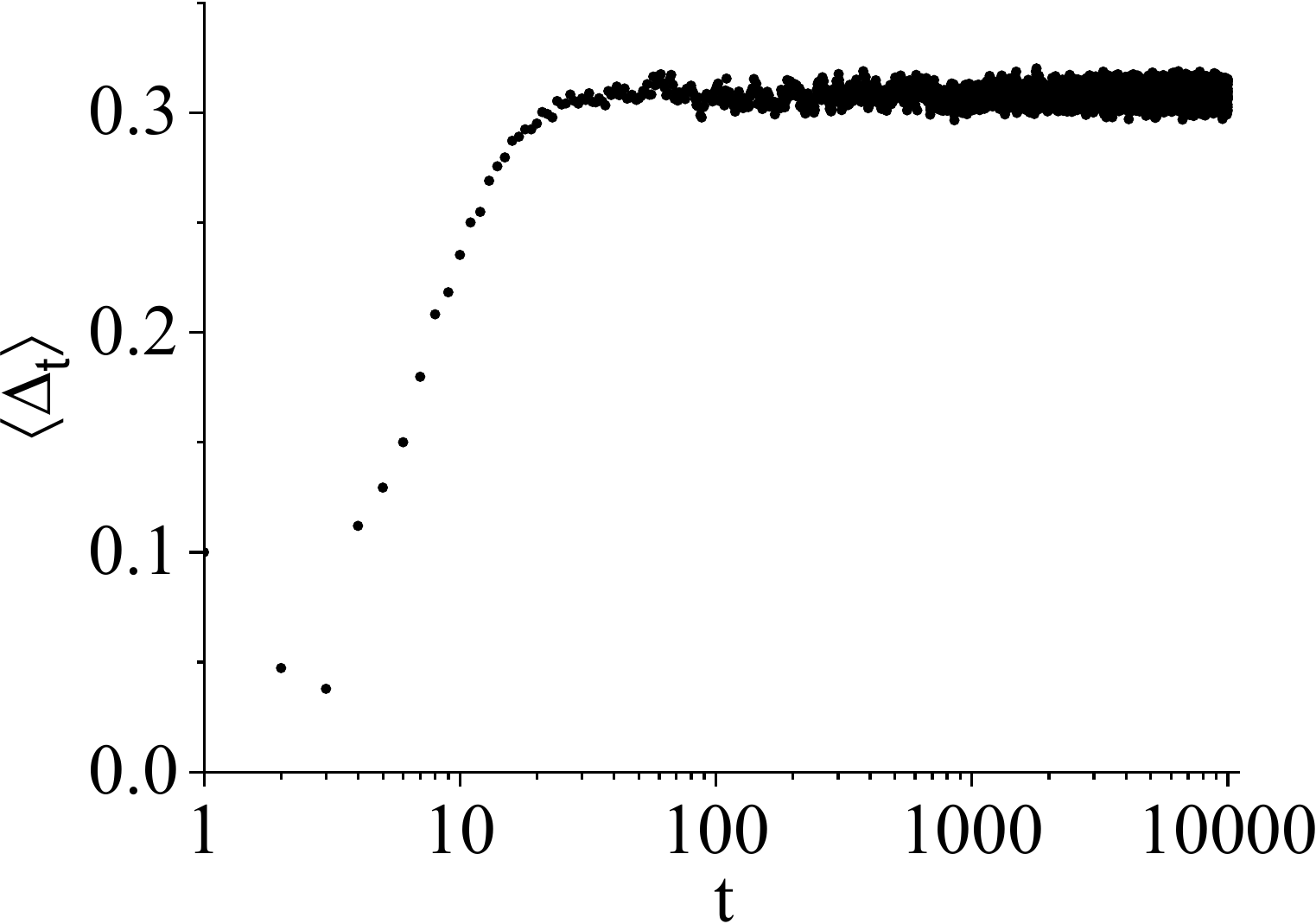}
\caption{}
\end{subfigure}
\begin{subfigure}[b]{0.7\linewidth}
\includegraphics[width=\linewidth]{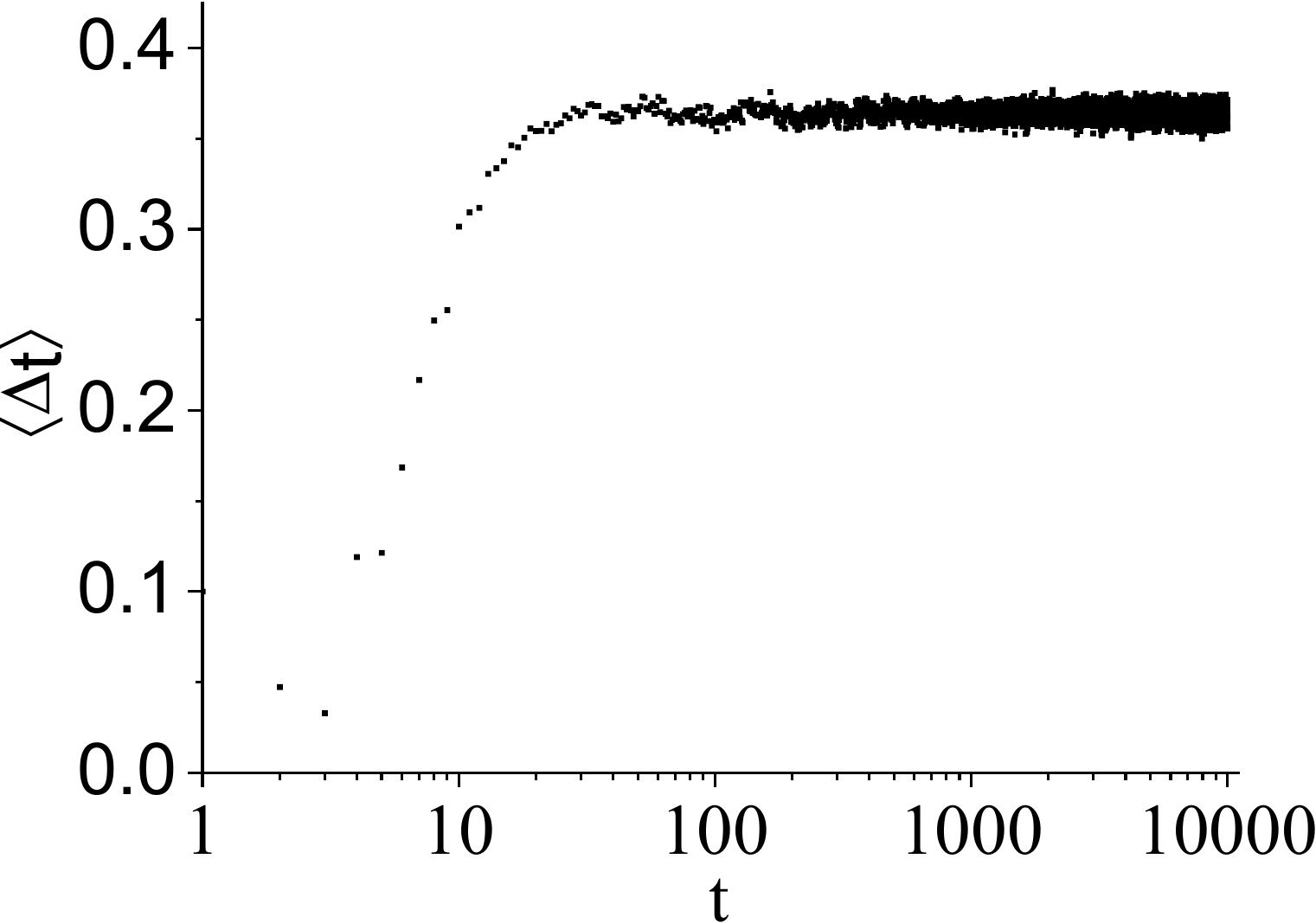}
\caption{}
\end{subfigure}
\caption{TM method results: $\langle\Delta_{t}\rangle$ against time for (a) uniform distribution and (b) symmetric triangular distribution having the range between $q_1 = 2.0$ and $q_2 = 2.53$. For black curve, initial $\Delta_0$ is $10^{-4}$. For red curve, $\Delta_0$ is $10^{-3}$. For blue and green curves, the values of $\Delta_0$ are $10^{-2}$ and $10^{-1}$, respectively. (Color Online.)}
\label{FIG:8}
\end{figure}
Fig.~\ref{FIG:8} illustrates this asymptotic behavior of $\langle\Delta_{t}\rangle$ as the number of iterations increases, for both distributions within the range $q_1 = 2.0$ and $q_2 = 3.5$. The ensemble averages of $\Delta_{t}$ are plotted in the chaotic regime next for different initial separations of $\Delta_t$ within the region where their values alternate between zero and non-zero with time. Remarkably, it is observed that the ensemble averages of $\Delta_t$ for any initial separations reach a steady state non-zero value after a finite number of iterations, which we denote as $\Delta_{\rm sat} = \langle\Delta_{\rm t\to \infty}\rangle$.

So, the two curves in Fig.~\ref{FIG:8} demonstrate that even though the separation between the two state variables oscillates between zero and non-zero values, the ensemble average of separation for both distributions ultimately saturates to specific values independent of initial $\Delta_t$. The saturation values, $\Delta_{\rm sat}$ for uniform and symmetric triangular distributions are about $0.3$ and $0.4$, respectively.

\begin{figure}[tbh]
\centering
\begin{subfigure}[b]{0.7\linewidth}
\includegraphics[width=\linewidth]{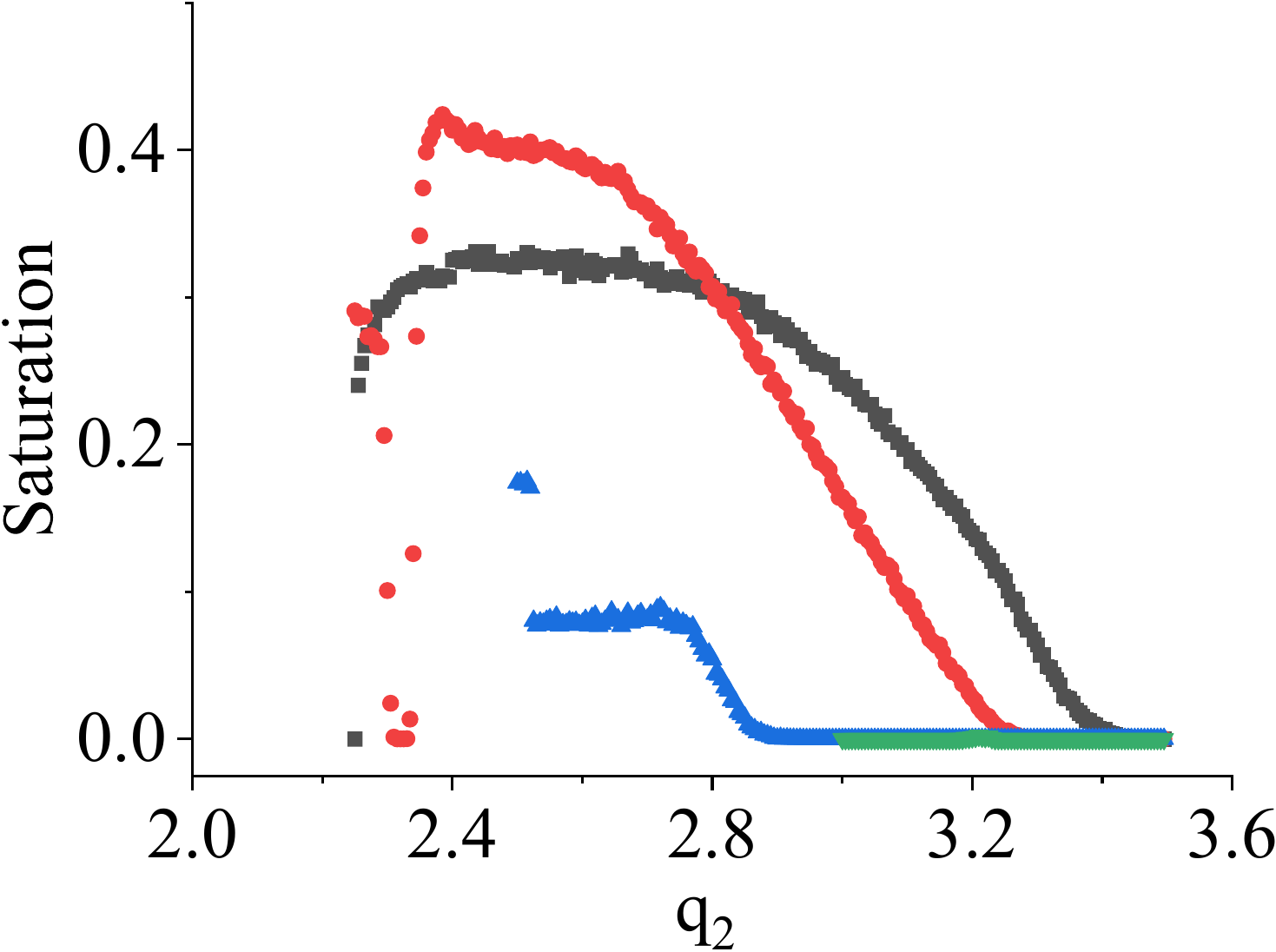}
\caption{}
\end{subfigure}
\begin{subfigure}[b]{0.7\linewidth}
\includegraphics[width=\linewidth]{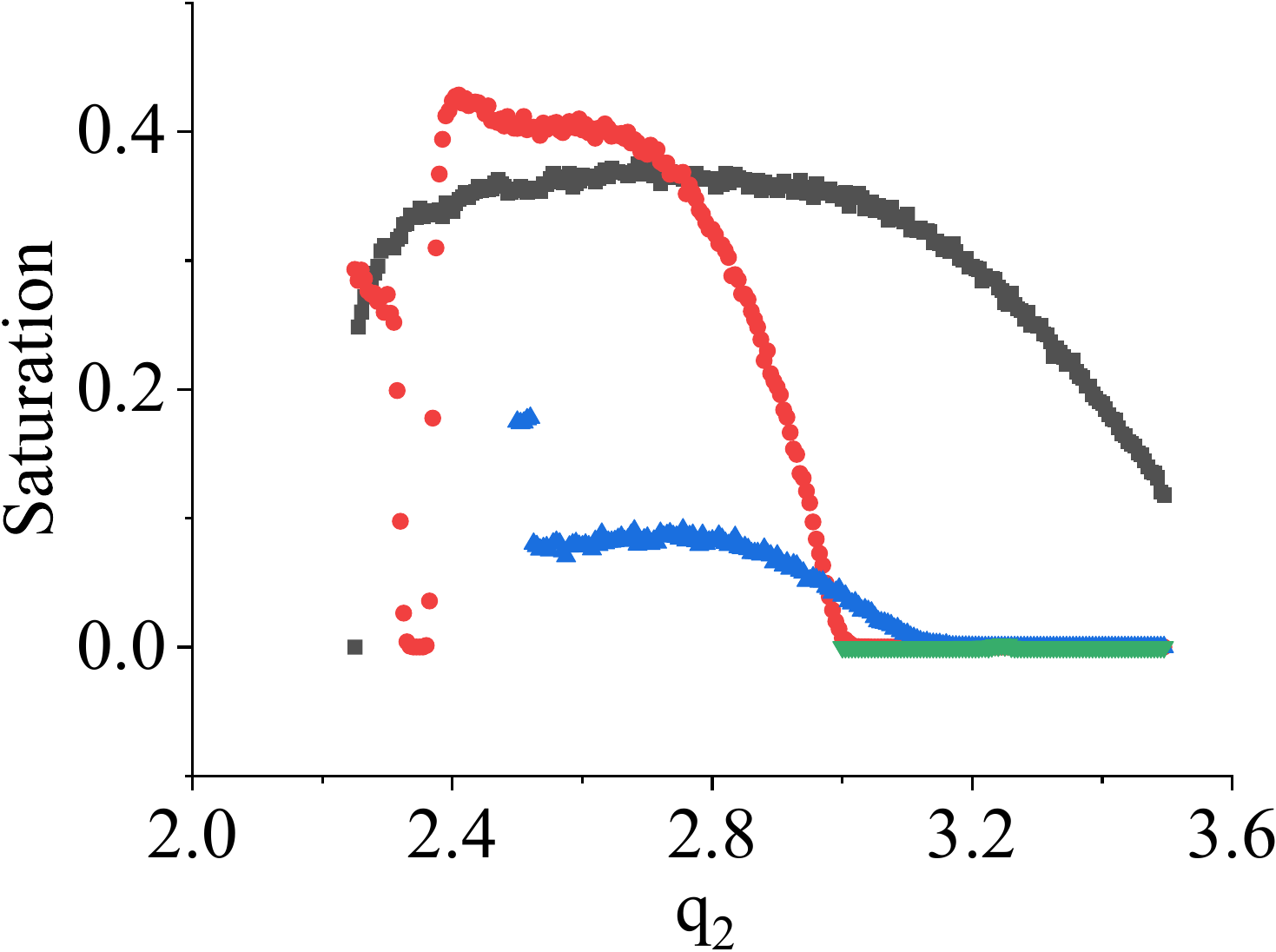}
\caption{}
\end{subfigure}
\caption{TM method results: The variation of saturation values of damage $\Delta_{\rm sat}$ with $q_2$ having $q_1$ is fixed at different values in the case of (a) uniform distribution and (b) symmetric triangular distribution. $q_2$ is varied in between $q_1$ and $3.5$. The different color denotes the different fixed values of $q_1$. The black curve denotes when $q_1 = 2.0$, the red curve denotes for $q_1 = 2.25$, the blue curve denotes for $q_1 = 2.5$, and the green curve corresponds to $q_1 = 3.0$. (Color online.)}
\label{FIG:9}
\end{figure}
Fig.~\ref{FIG:9}(a) and Fig.~\ref{FIG:9}(b) depict the saturation values of damage, denoted by $\Delta_{\rm sat}$, with the variation of $q_2$ for both uniform and symmetric triangular distributions, while maintaining $q_1$ at various fixed values. The specific values of $q_1$ are specified in the caption of Fig.~\ref{FIG:9}. It is evident from both figures that $\Delta_{\rm sat}$ initially maintains a non-zero value and then decreases with increasing values of $q_2$, while $q_1$ is held constant. Eventually, $\Delta_{\rm sat}$ reaches zero after surpassing a certain threshold value of $q_2$. This observation suggests that the chaotic behavior of the system transforms into periodic behavior when $\Delta_{\rm sat}$ reaches zero from a non-zero value after a specific $q_2$ value for uniform and symmetric triangular distributions. However, it is worth noting that the threshold value of $q_2$ for a fixed $q_1$ is distinct for each distribution. An essential finding from both graphs is that as $q_1$ increases, $\Delta_{\rm sat}$ approaches zero more rapidly, and when $q_1 = 3.0$, $\Delta_{\rm sat}$ is always zero irrespective of $q_2$. This implies that reducing the separation between $q_1$ and $q_2$ by approaching $q_1$ towards $q_2$ leads to a faster transition of the random map into a non-random state.

\begin{figure}[tbh]
\centering
\begin{subfigure}[b]{0.7\linewidth}
\includegraphics[width=\linewidth]{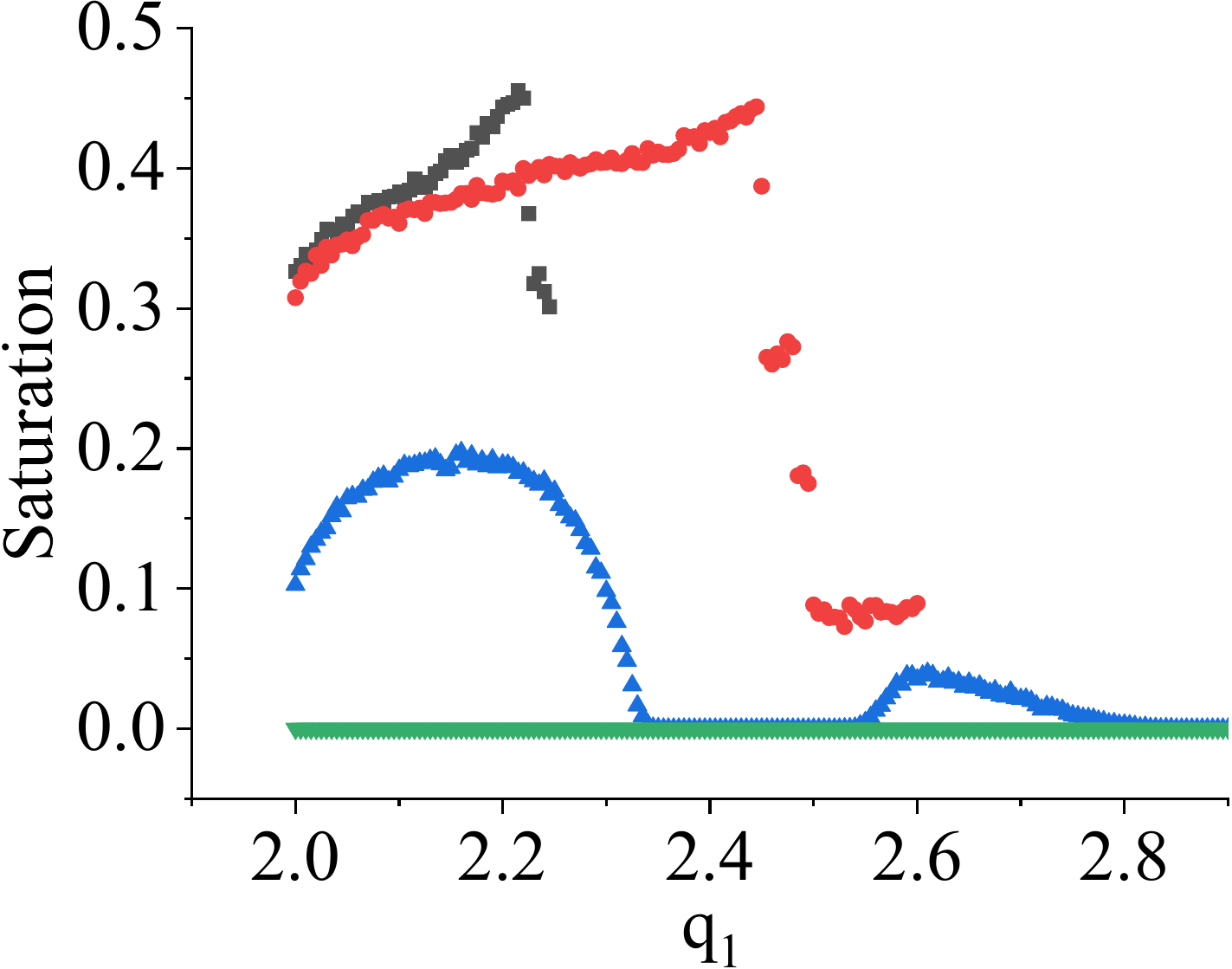}
\caption{}
\end{subfigure}
\begin{subfigure}[b]{0.7\linewidth}
\includegraphics[width=\linewidth]{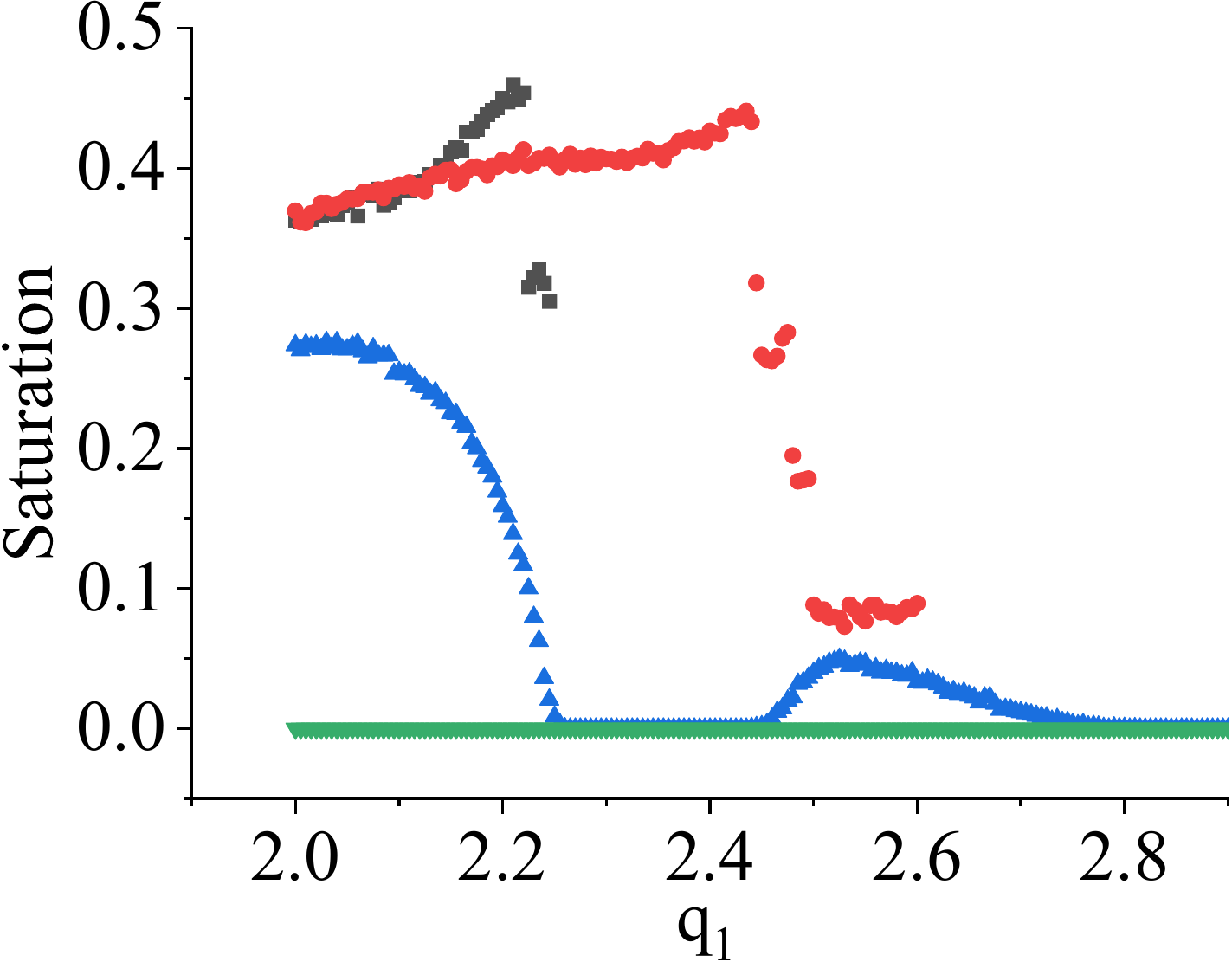}
\caption{}
\end{subfigure}
\caption{TM method results: The variation of saturation values of damage $\Delta_{\rm sat}$ with $q_1$ having $q_2$ is fixed at different values in the case of (a) uniform distribution and (b) symmetric triangular distribution. $q_1$ is varied in between $2.0$ and $q_2$. The different color denotes the different fixed values of $q_2$. The black curve denotes when $q_2 = 2.25$, the red curve denotes for $q_2 = 2.5$, the blue curve denotes for $q_2 = 3.0$, and the green curve corresponds to $q_2 = 3.5$. (Color online.)}
\label{FIG:9a}
\end{figure}
Moreover, in Fig.~\ref{FIG:9a}(a) and Fig.~\ref{FIG:9a}(b), for both distributions, when $q_2$ is fixed at various values, and $q_1$ is varied within the range of 2.0 to $q_2$, the parameter $\Delta_{\rm sat}$ exhibits irregular patterns. In the case where $q_2 = 3.5$, $\Delta_{\rm sat}$ consistently remains at zero for all values of $q_1$, indicating a system-wide periodic behavior across the parameter space. However, for other $q_2$ values, $\Delta_{\rm sat}$ switches between zero and non-zero values as $q_1$ gradually approaches the $q_2$ values. This toggling effect lacks periodicity with respect to $q_1$. Consequently, the stochastic map demonstrates a hopping phenomenon between periodic and chaotic waveforms with variations in $q_1$.

\section{Computation of $\langle\Delta_{\mathrm{sat}}\rangle$ for TM method}
\label{analytical_treatment}
In the case of the traditional method (TM), the stochastic map $g$ is applied for the initial value $v_{\rm C0}$ and its perturbation value ($v_{\rm C0} + \Delta$) with the same random parameter chosen from identical distribution for each iteration. After $n$ number of iterations, the values of the state variable and its perturbation will be mapped as $g^n(v_{C0})$ and $g^n(v_{C0} + \Delta)$. The ensemble average of absolute distance between the two values of the state variables will be defined as $\langle\Delta_{\mathrm{sat}}\rangle$. The ensemble average has to be taken against the time-invariant probability distribution of $p_t(v_C)$. As $\langle\Delta_{\mathrm{sat}}\rangle$ is the ensemble average of the absolute distance of $g^n(v_{C0})$ and $g^n(v_{C0} + \Delta)$, the functional form $\langle\Delta_{\mathrm{sat}}\rangle$ in terms of root mean square average \cite{KHALEQUE2014599} can be written as
\begin{equation}\label{eq_5}
\langle\Delta_{\mathrm{sat}}\rangle=\sqrt{\langle(g^n(v_{C0}+\Delta)-g^n(v_{C0}))^{2}\rangle} 
\end{equation}

The function is analyzed $\langle(g^n(v_{c0}+\Delta)$ - $g^n(v_{c0}))^{2}\rangle$ to check its behavior under different conditions.
\begin{equation}\label{eq_6}
\begin{aligned}
\langle(g^n(v_{C0}+\Delta)-g^n(v_{C0}))^{2}\rangle =& \langle((g^n(v_{C0}+\Delta)^2+\\
&(g^n(v_{C0})^2-g^n(v_{C0}+\Delta)g^n(v_{C0}))\rangle \\
   =& \langle(g^n(v_{C0}+\Delta)^2\rangle+\langle(g^n(v_{C0})^2\rangle\\
   &\langle-2g^n(v_{C0}+\Delta)g^n(v_{C0})\rangle
\end{aligned}
\end{equation}

Let us discuss these three terms individually. From the first term:
\begin{equation}
\begin{aligned}
   \langle(g^n(v_{C0}+\Delta)^2\rangle=\int  (g^n(v_{C0}+\Delta)^2 p(v_C) dv_C
\end{aligned}
\end{equation}

We know that as $p(v_C)$ would be same as $p(v_C+\Delta)$ because as $t\rightarrow\infty$, $p_t(v_C)$ and $p_t(v_C+\Delta)$ would achieve the same stationary form  $p(v_C)$.

We can also write the above equation as 
\begin{equation}
\begin{aligned}
   \langle(g^n(v_{C0}+\Delta)^2\rangle=&\int  (g^n(v_{C0}+\Delta)^2 p(v_C+\Delta) dv_C\\
    = & \int  (g^n(v_{C0}+\Delta)^2 p(v_C+\Delta) d(v_C+\Delta)\\
    = & \int  (g^n(z)^2 p(z) d(z)\\
\end{aligned}
\end{equation}
Which is the same as the second term. 

From the third term, we can write,
\begin{equation}
\begin{aligned}
\langle-2g^n(v_{C0}+\Delta)g^n(v_{C0})\rangle=-2\langle g^n(v_{C0}+\Delta)g^n(v_{C0})\rangle
\end{aligned}
\end{equation}

There are two different cases for the third term.
\begin{itemize}
\item[1.]  Case~$1$: assuming $g^n(v_{C0}+\Delta)\stackrel{n\rightarrow\infty}{=}g^n(v_{C0})$ (Non-chaotic regime), then, it can be written as
\begin{equation}
\begin{aligned}
\langle-2g^n(v_{C0}+\Delta)g^n(v_{C0})\rangle=-2\langle
g^n(v_{C0})^2\rangle
\end{aligned}
\end{equation}
If we put this expression in Equation~(\ref{eq_6}), we can see that $\Delta_{\mathrm{sat}}$ is zero.

\item[2.] Case~$2$: $g^n(v_{C0}+\Delta)\stackrel{n\rightarrow\infty}{\neq}g^n(v_{C0})$ (chaotic regime) and $\Delta_{\mathrm{sat}}$ cannot be $0$.
\end{itemize}

\section{Results for NVN method}
\label{results_nvn}

As previously stated, for the NVN (Nature versus Nurture) method, the process commences by considering two initially identical values of $v_{\rm C}(t)$, which then evolve independently. These two evolutions follow distinct sets of random control parameters drawn from identical distributions. The selection of control parameters is made from both uniform and symmetric triangular distributions.

\subsection{Non-Chaotic regime}
\label{nonchaoticregime}
In the NVN method, due to the utilization of identical initial conditions in conjunction with distinct randomly chosen bifurcation parameters for each iteration, it is observed that the average separation $\langle\Delta_{t}\rangle$ between the state-variable and its perturbation cannot reach zero. This condition prevents the merging of the state variable and its perturbation, which implies the absence of periodic behavior. Consequently, the absence of a non-chaotic regime is observed in the system. The inherent nature of the NVN method, with its specific parameter selection strategy, results in the persistence of chaotic dynamics throughout the simulation.

\subsection{Chaotic regime}
\label{chaotic regime}
\begin{figure}[tbh]
\centering
\begin{subfigure}[b]{0.7\linewidth}
\includegraphics[width=\linewidth]{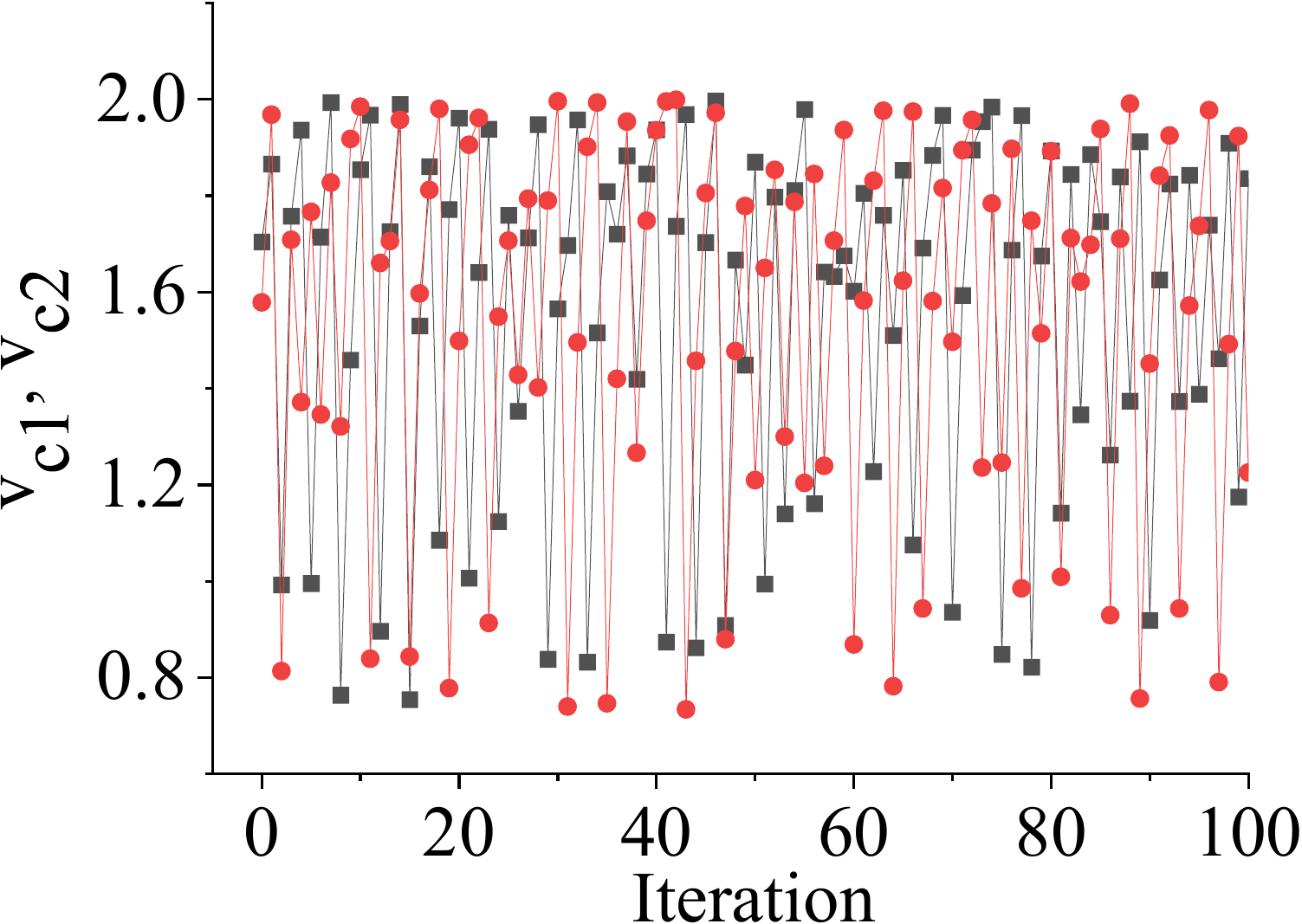}
\caption{}
\end{subfigure}
\begin{subfigure}[b]{0.7\linewidth}
\includegraphics[width=\linewidth]{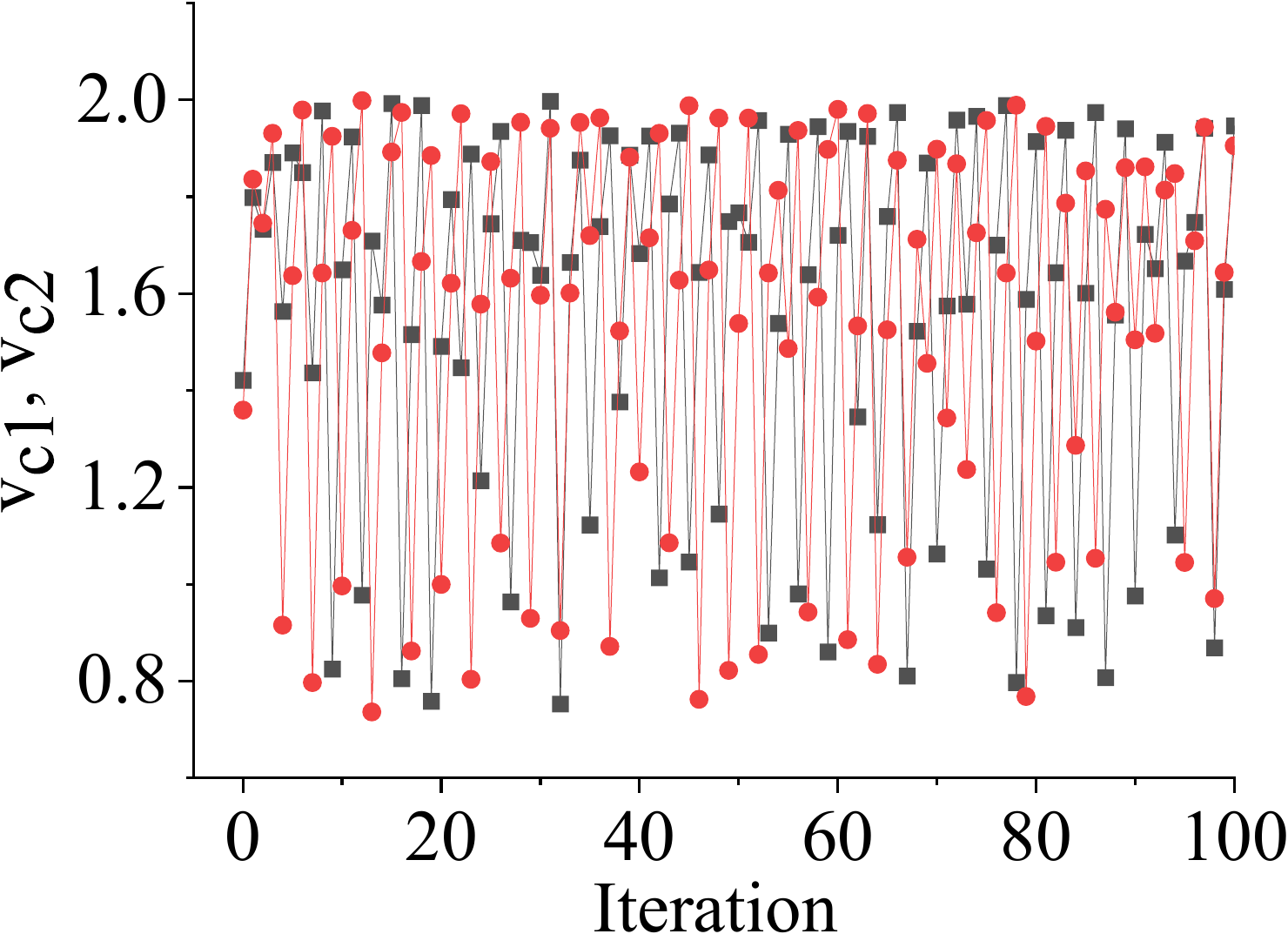}
\caption{}
\end{subfigure}
\caption{NVN method results: $ v_{\rm C1}$ (black in color) and $v_{\rm C2}$ (red in color) against time (a) for uniform distribution and (b) for symmetric triangular distribution having the range between $2.0$ and $2.53$. (Color online.)}
\label{fig:chaosnvnuniform22}
\end{figure}
In Fig.~\ref{fig:chaosnvnuniform22}, we can observe the evolutions of two initially identical state variables over time using the NVN method for both distributions. The two figures clearly demonstrate that the state variables do not merge as time progresses. Consequently, the ensemble average $\langle\Delta_{t}\rangle$ of the state variables never reaches zero but tends to approach a constant value with time. These average values remain consistent for both distributions, similar to the trend observed in the TM (Traditional Method) case.

This behavior further reinforces that the NVN method sustains chaotic dynamics throughout its evolution, regardless of the chosen distribution. The inability of the state variables to merge and the persistent non-zero average separation signifies the continuous presence of chaotic behavior without any periodic regimes in the system.

\begin{figure}[tbh]
\centering
\begin{subfigure}[b]{0.7\linewidth}
\includegraphics[width=\linewidth]{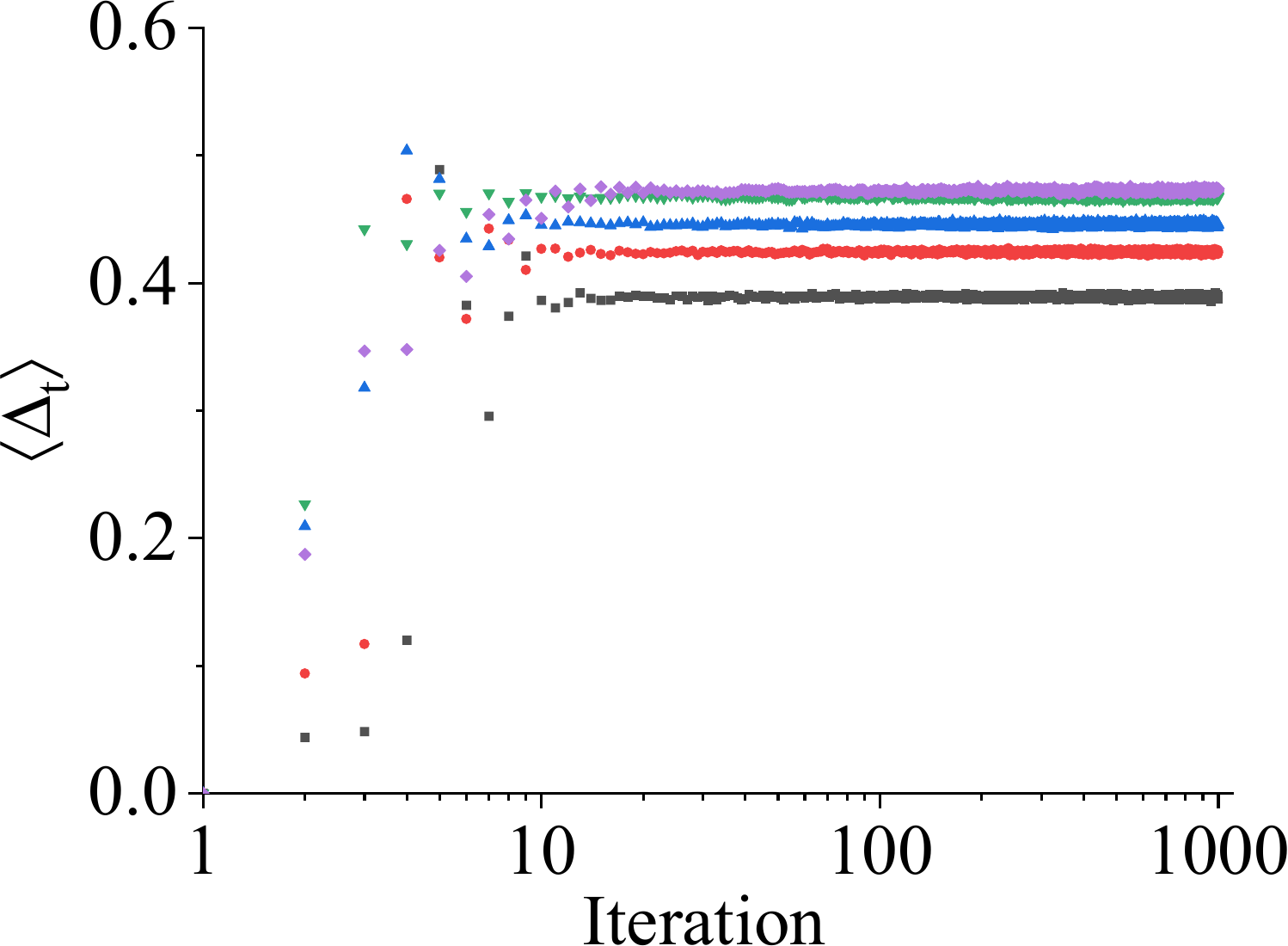}
\caption{}
\end{subfigure}
\begin{subfigure}[b]{0.7\linewidth}
\includegraphics[width=\linewidth]{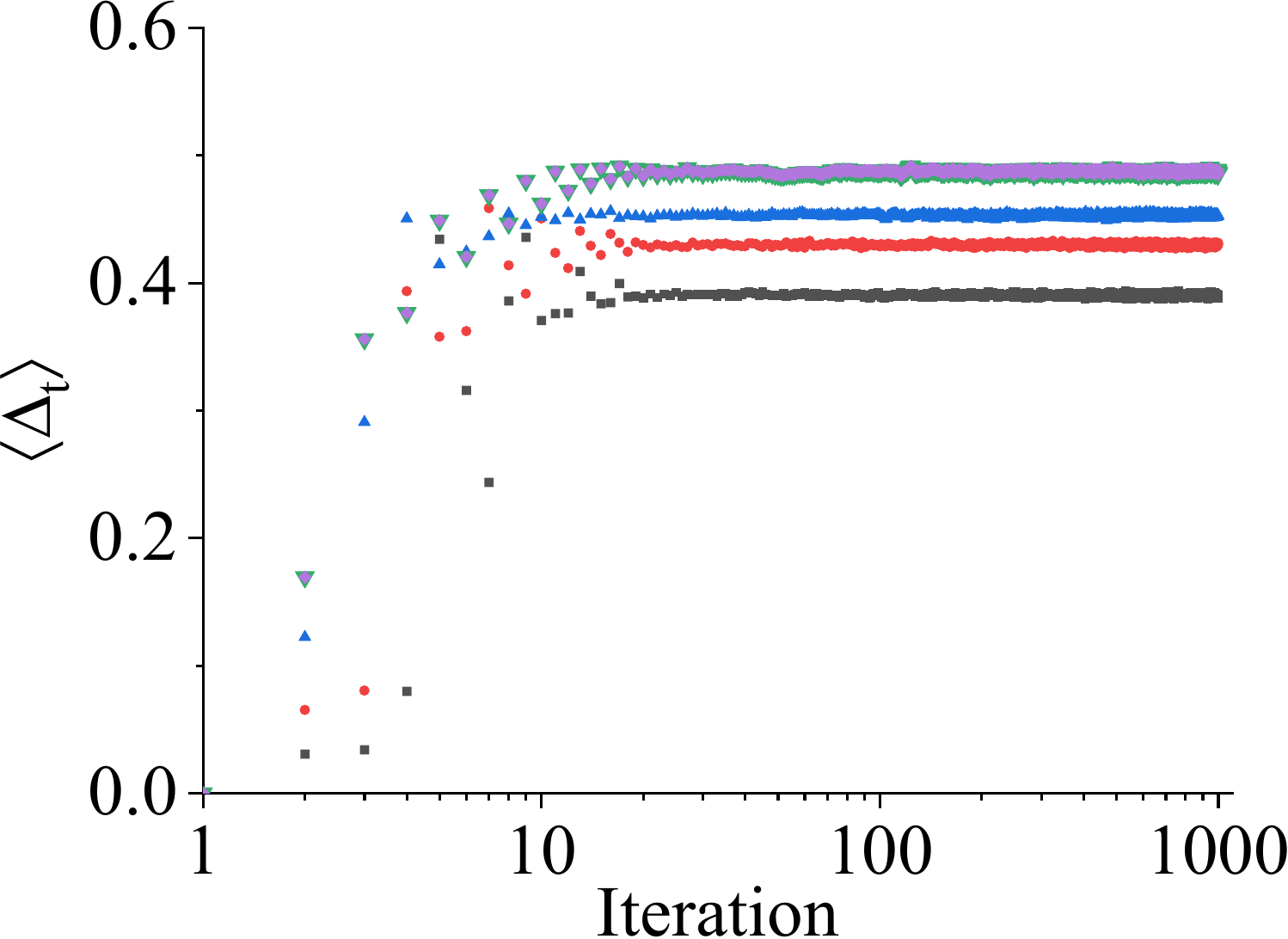}
\caption{}
\end{subfigure}
\caption{NVN method results: The time evolution of damage, $\Delta_{t}$, for various $q_1$ and $q_2$ values in the case of (a) uniform distribution, and (b) symmetric triangular distribution. For each of the figures, the pink color denotes $q_1 = 2.25$ and $q_2 = 3.5$, green color denotes $q_1 = 2.0$ and $q_2 = 3.5$, blue color denotes $q_1 = 2.0$ and $q_2 = 3.0$, red color denotes $q_1 = 2.0$ and $q_2 = 2.53$, and black color denotes $q_1 = 2.0$ and $q_2 = 2.25$. (Color online.)}
\label{fig:chaosnvnuniform}
\end{figure}
In Fig.~\ref{fig:chaosnvnuniform}, the values of damage $\Delta_{\rm t}$ are presented over time for two distributions with different ranges from where the random parameters are chosen using the NVN method. The damage values corresponding to various combinations of $q_1$ and $q_2$ initially undergo changes from their initial values and then settle into constant values, i.e., they saturate. These saturation values of $\Delta_{\rm t}$ remain non-zero and differ between the parameter ranges of the two distributions. However, it is noteworthy that despite the differences in parameter ranges, the saturation values are identical for both distributions in the same parameter ranges, as seen in Fig.~\ref{fig:chaosnvnuniform}. The persistent non-zero saturation of $\Delta_{\rm t}$ reinforces the presence of chaotic behavior in the system, as the damage values never tend to zero but reach stable non-zero values for both distributions. This observation aligns with the absence of a non-chaotic regime in the NVN method, as the chaotic dynamics dominate the system's behavior irrespective of the parameter distributions.

\begin{figure}[tbh]
\centering
\begin{subfigure}[b]{0.7\linewidth}
\includegraphics[width=\linewidth]{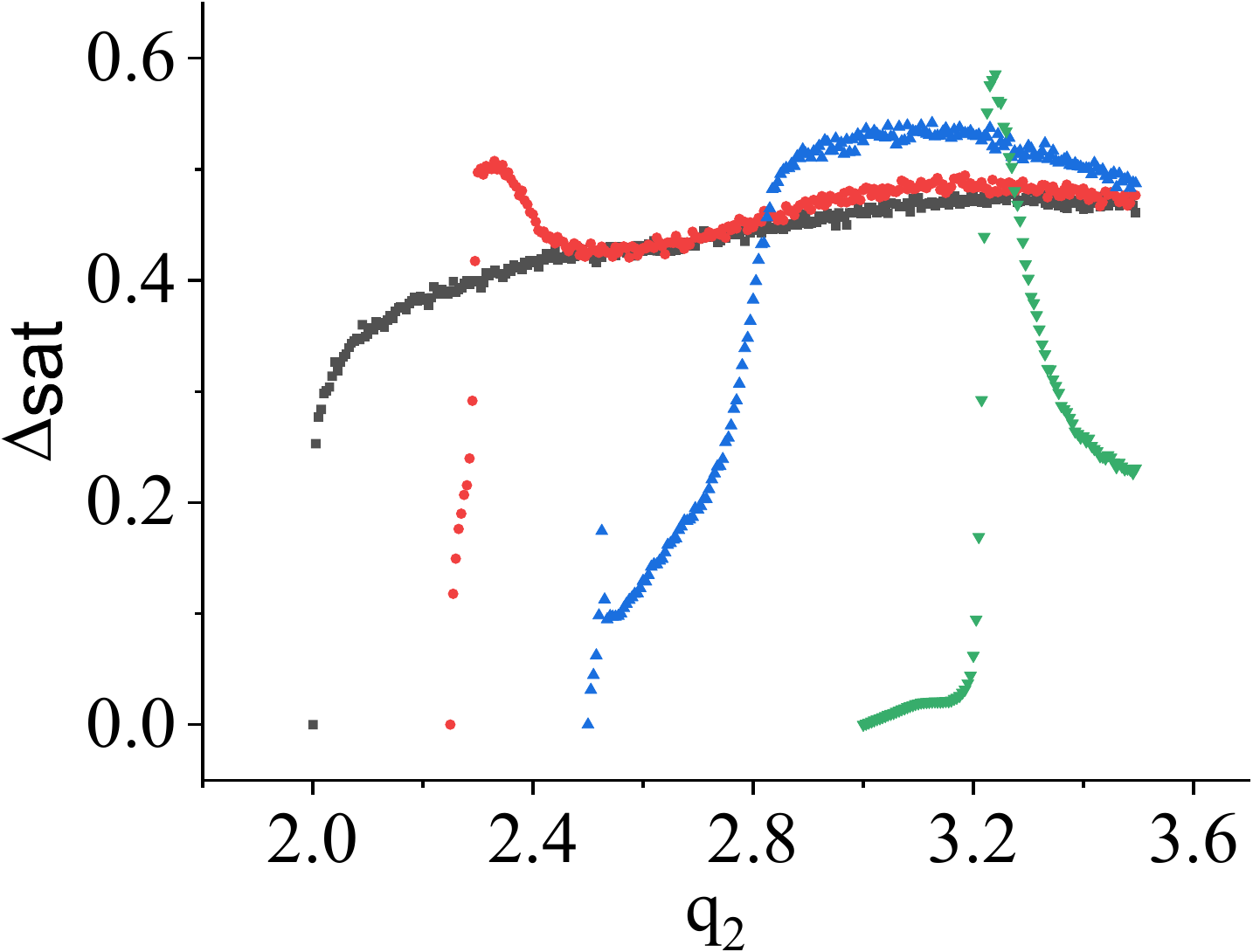}
\caption{}
\end{subfigure}
\begin{subfigure}[b]{0.7\linewidth}
\includegraphics[width=\linewidth]{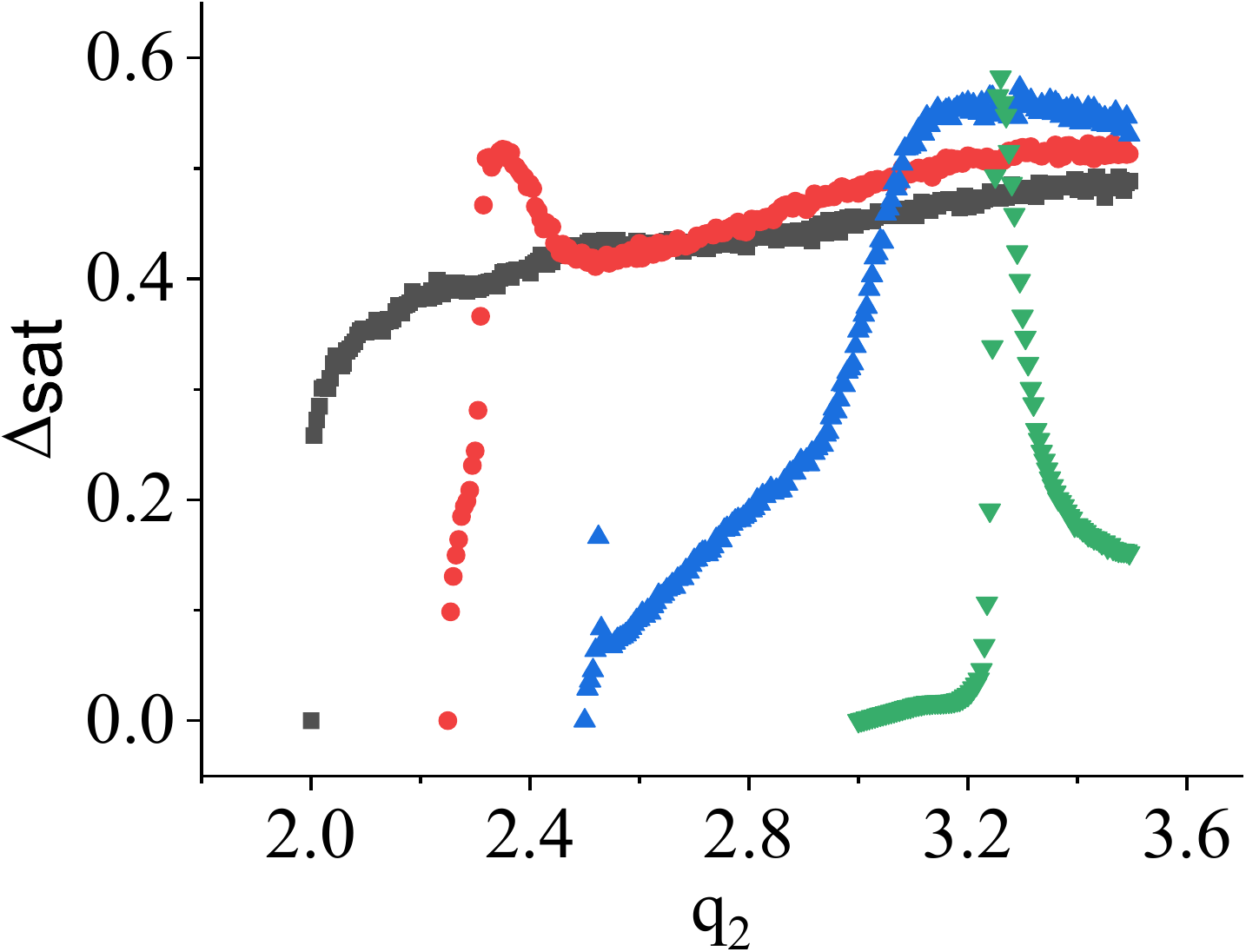}
\caption{}
\end{subfigure}
\caption{NVN method results: The saturation values of damage, $\Delta_{\rm sat}$, with the variation of $q_2$ for different fixed values of $q_1$ in the case of (a) uniform distribution and (b) symmetric triangular distribution. The curves having different colors denote the different values of $q_1$. $q_2$ is varied in between $q_1$ and $3.5$. The black curve denotes $q_1=2.0$, the red curve denotes $q_1=2.25$, the blue curve denotes $q_1 = 2.5$, and the green denotes $q_1 = 3.0$. (Color online.)} 
\label{FIG:10}
\end{figure}

In Fig.~\ref{FIG:10}, the variations of the saturation value of damage, denoted as $\Delta_{\rm sat}$, are depicted with respect to the parameter $q_2$, while $q_1$ is kept fixed at different values for both uniform and symmetric triangular distributions. The figures illustrate that as the value of $q_2$ increases from $q_1$ towards $3.5$, $\Delta_{\rm sat}$ increases from zero. This observation indicates that for each fixed value of $q_1$, the random map transitions from a periodic behavior to a chaotic regime.

An interesting behavior is observed when the fixed value of $q_1$ is set to $2.0$. In this case, $\Delta_{\rm sat}$ increases from zero and eventually stabilizes at a fixed value close to $0.4$ as $q_2$ is iterated forward. However, as the fixed value of $q_1$ is increased from $2.0$, the variations of $\Delta_{\rm sat}$ first reach maximum values from zero as $q_2$ varies and then start to decrease. Each $\Delta_{\rm sat}$ has a distinct peak value before it begins to decrease in different fashions. This observation implies that as the fixed value of $q_1$ changes, the variations of $\Delta_{\rm sat}$ exhibit abrupt changes with respect to $q_2$, and there is no apparent correlation between the curves.

In the context of the TM method, as shown in Fig.~\ref{FIG:9}, under the specified condition, $\Delta_{\rm sat}$ exhibited a state of constancy initially until reaching a certain threshold of $q_2$, after which it progressively diminished to zero with further increments in $q_2$. At $q_1 = 3.0$, $\Delta_{\rm sat}$ is always zero for all the values of $q_2$. Conversely, when applying the NVN method, $\Delta_{\rm sat}$ displayed entirely unpredictable and irregular patterns of behavior under the same condition.

\begin{figure}[tbh]
\centering
\begin{subfigure}[b]{0.7\linewidth}
\includegraphics[width=\linewidth]{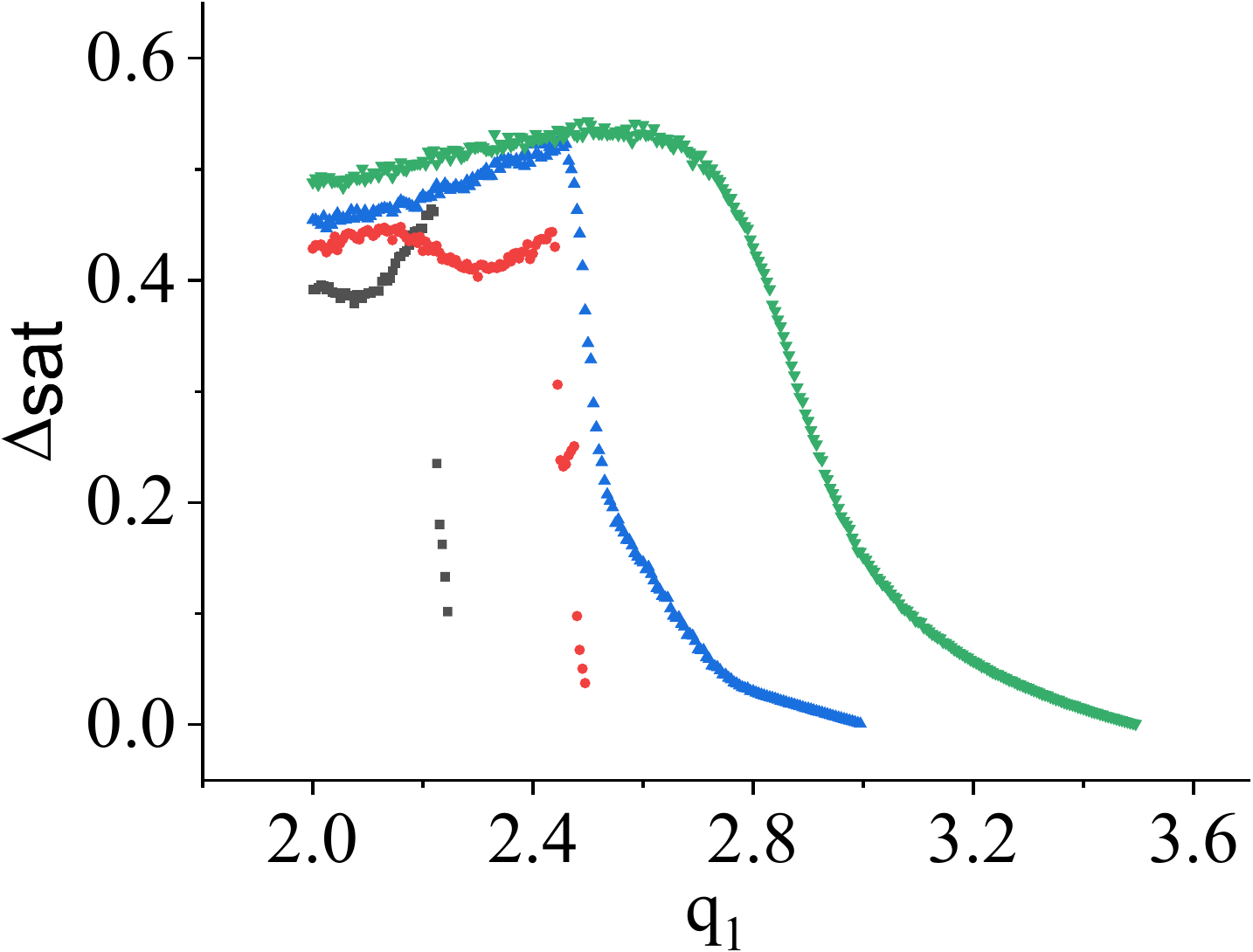}
\caption{}
\end{subfigure}
\begin{subfigure}[b]{0.7\linewidth}
\includegraphics[width=\linewidth]{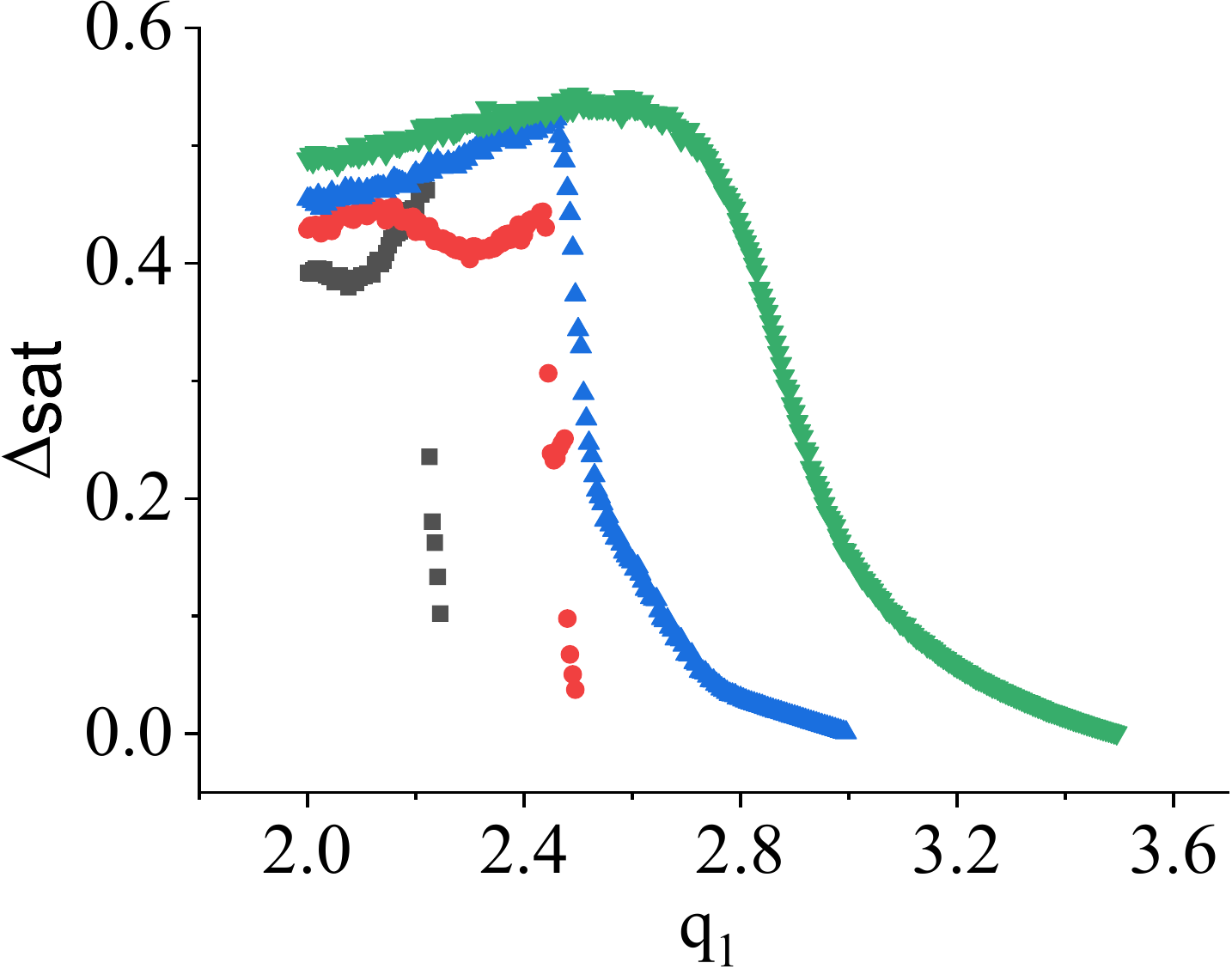}
\caption{}
\end{subfigure}
\caption{NVN method results: The variations of the saturation values of damage $\Delta_{\rm sat}$ with $q_1$ when $q_2$ is fixed at different values for (a) uniform distribution (b) symmetric triangular distribution. Different colors in each figure denote different fixed values of $q_2$. $q_1$ is varied in between $2.0$ and $q_2$. The black color denotes the variation of $\Delta_{\rm sat}$ for $q_2=2.25$, the red color is for $q_2=2.5$, the blue color is for $q_2 = 3.0$, and the green color is for $q_2 = 3.5$. (Color online.)}
\label{FIG:delta_sat_q_variation}
\end{figure}
The variations of the saturation of damage $\Delta_{\rm sat}$ with $q_1$ by fixing $q_2$ at different fixed values for uniform and symmetric triangular distributions are plotted next, as shown in Fig.~\ref{FIG:delta_sat_q_variation}(a) and Fig.~\ref{FIG:delta_sat_q_variation}(b), respectively. Here, for all the curves, $\Delta_{\rm sat}$ have certain non-zero values up to certain ranges and then attenuate towards zero values while $q_1$ varies in the positive direction. This implies that in this condition, when $q_1$ is varied, and $q_2$ is fixed to a particular value, the random map shows a transition to an ordered behavior from the chaotic ones. While $q_2$ is chosen at different fixed values, with the increase of $q_2$, $q_1$ takes more value to attenuate $\Delta_{\rm sat}$ towards zero values.

In contrast to the NVN method, when the same conditions are applied to the TM method, as shown in Fig.~\ref{FIG:9a}, $\Delta_{\rm sat}$ remains consistently at zero at $q_2 = 3.5$, but erratic behaviors for other $q_2$ values, suggesting a significant difference in behavior between the two methods. Specifically, the NVN method exhibits complex and chaotic dynamics changes to a periodic behavior with varying values of $q_1$. In contrast, the TM method maintains zero and non-zero transitions of $\Delta_{\rm sat}$ values under the same condition.

The stochastic, random, non-smooth one-dimensional map exhibits distinct dynamical stabilities in the saturation value of damage ($\Delta_{\rm sat}$) when various types of probability distributions are employed. This analogous non-conformity in behavior for two dissimilar distributions was similarly observed in the context of a random continuous map, as demonstrated in the case of a logistic map \cite{doi:10.1142/S0129183115500862}.

\section{Conclusions}
\label{conclusions}
This paper examines the impact of randomness on a piecewise smooth map~(\ref{eq_1}) derived from an inductorless circuit, which is representative of a wide range of realistic systems. By randomly selecting the bifurcation parameter $V_1$ from two distributions, we observe various behaviors of the state variable $v_{\rm C}(t)$. Depending on the parameter distribution ranges and initial conditions, the system exhibits fully ergodic or semi-ergodic dynamics. Notably, despite the presence of chaotic waveforms in $v_{\rm C}(t)$, the ensemble average of the variable converges to a constant saturation value over time.

We analyze the probability density function of the map within a parameter range randomly chosen from a distribution characterized by $q_1=2.0$ and $q_2=3.5$. This specific parameter region demonstrates the phenomenon of reverse period incrementing cascade bifurcation observed in the case of the nonrandom map. This curve signifies that the probability of finding the state variable is mostly on the switching manifold. Importantly, the probability density function remains invariant regardless of the chosen initial conditions within the fixed bifurcation parameter range. 

Regarding the deterministic regime, we anticipate chaotic behavior after passing through various periodic attractors in parameter space \cite{PhysRevLett.65.2935, strogatz2018nonlinear}. Interestingly, our selected map showcases an interplay between periodic orbits and chaos under specific parameter settings. We find that the introduction of randomness in the switching circuit leads to nonchaotic behavior ($\Delta_t$ tending to zero as time progresses) within the parameter space for both distributions. Furthermore, periodic behavior emerges in the parameter space in the stochastic regime, even though the nonrandom map exhibits chaos within the same range. In the nonchaotic regime, the separations between the maximum and minimum values of the state variable follow the relationship $\langle\Delta_t\rangle \propto {\Delta_0^\alpha} \exp (\lambda t)$, which differs from the behavior of the nonrandom map. At that instant, we determine the Lyapunov exponent ($\lambda$) and observe its unconventional dependence on the distribution's asymmetry.

Additionally, we investigate the TM and NVN methods and examine the transitions from chaos to periodic behavior within the parameter space defined by the ranges of $q_1$ and $q_2$. The traditional method exhibits periodic behavior from chaos around $q_2 \approx 3.2$ when $q_1$ is fixed at $2.0$. In contrast, chaos does not occur when $q_2$ is fixed and $q_1$ is varied, suggesting a different behavior. Notably, the NVN method reveals a transition from chaotic to periodic behavior when $q_2$ is fixed, and $q_1$ varies. Surprisingly, chaos emerges instead of transitioning to periodic behavior when $q_1$ is fixed and $q_2$ is varied. For different fixed values of $q_1$, the variations of $\Delta_{\rm sat}$ with $q_2$ do not follow any pattern. 

In the traditional method, the separation $\Delta_t$ oscillates between zero and nonzero values in a chaotic regime over time, indicating varying separations between two state variables. However, attempts to establish correlations with the ratios of consecutive zero and nonzero separations yield random results, unlike the convergence to a fixed value observed in the nonrandom case. This proves that the system shows random behavior due to the randomness in the parameter values.

Our findings also confirm findings of previous studies suggesting the independent nature of the TM and NVN methods \cite{PhysRevE.88.040101, KHALEQUE2014599}. This work highlights the determination of saturation values for damage under varying $q_1$ and $q_2$, confirming the consistency of these methods. Notably, we observe a chaotic regime in the NVN method when $q_1$ and $q_2$ differ, reminiscent of the phenomenon of damage spreading in the opinion dynamics model \cite{KHALEQUE2014599}. 

The chaos generating circuit considered in this study proves advantageous for practical implementations, particularly within the realms of chaos communication and experimental investigation of border collision bifurcation phenomena in an electronic circuit. In the context of physical applications, meticulous determination of parameter values poses a considerable challenge. There are also ripples in addition to the precise values of parameters. Consequently, this research holds significant utility for the practical utilization of this piecewise smooth map in terms of the expected dynamical responses.

In scenarios involving a non-random, non-smooth map, one may encounter various unconventional bifurcation behaviors that are not observable in continuous maps. Certain bifurcations within this context may give rise to potentially hazardous situations in physical systems that entail specific switching conditions. this work also presents a period incrementing cascade featuring chaotic inclusions amidst periodic attractors. Contrasting this, we observe period-doubling bifurcations in a Logistic map, which progressively transition into chaotic attractors. Notably, there exist periodic windows interspersed amid chaotic orbits. 
In future, this work can be extended to investigate the dynamics of a different class of non-smooth map characterized by alternative non-smooth bifurcations and subsequently compare the observed dynamics with those obtained from the map employed in our current study.

\section*{Declarations}
The authors would like to acknowledge Enterprise Ireland SEMPRE project DT-2020-0243-A, Science Foundation Ireland NexSys 21/SPP/3756, and Sustainable Energy Authority of Ireland TwinFarm RDD/604 project and Science Foundation Ireland TrAin project 22/NCF/FD/10995.

\section*{Conflict of interest}
The authors declare that they have no conflict of interest.

\section*{Availability of data}
Not applicable.

\section*{Availability of code}
All codes implemented in this paper is available upon request from the corresponding author.

\bibliographystyle{spmpsci}       

\end{document}